# Identifying compromise solutions across multiple treatment benefit–harm profiles in multivariate network meta-analysis

**Theodoros Evrenoglou[1], Anna Chaimani[2], Gerta Rücker[1], Guido Schwarzer[1]**

[1] *Institute of Medical Biometry and Statistics, Faculty of Medicine and Medical Center-University of Freiburg, Freiburg im Breisgau, Germany*

[2] *Oslo Center for Biostatistics and Epidemiology, Department of Biostatistics, University of Oslo, Oslo, Norway*

Corresponding author: Dr. Theodoros Evrenoglou

Institute of Medical Biometry and Statistics, Faculty of Medicine and Medical Center - University of Freiburg, Freiburg, Germany

email: theodoros.evrenoglou@uniklinik-freiburg.de

**Abstract**

Health technology assessment often aims to balance evidence across multiple treatments and outcomes and assess treatment benefit–harm profiles. Multivariate network meta-analysis (mvNMA) is well suited to this task, as it synthesizes evidence across treatments and outcomes. However, the potential of mvNMA has been overlooked, leading decision-makers to rely on outcome-specific NMAs that ignore outcome correlations. Based on independent NMA models, previous work proposed methods such as multivariate P-scores or spie charts to assess treatment benefit–harm profiles by yielding across-outcomes treatment hierarchies. In this article, we propose an mvNMA-based framework for assessing treatment benefit–harm profiles. Our framework uses a hybrid-correlation mvNMA model and incorporates DuMouchel priors for treatment effects, enabling estimation of all treatment effects even when outcomes are missing for some treatments. We then extend common NMA ranking metrics, including SUCRA, the probability of being best, and median ranks, to the mvNMA setting. Based on these metrics, we adapt the VIKOR algorithm for mvNMA. This deterministic multicriteria decision method quantifies overall and worst-case treatment performance and, by balancing these aspects, produces an across-outcomes ranking while identifying treatments offering the optimal compromise between these two aspects. To facilitate use of our framework, we developed the R package **mvnma**. We applied our framework to two clinical datasets: one comparing eleven physical therapy treatments across two outcomes and another comparing nine antidepressants across five outcomes. Our framework provides an alternative strategy for assessing treatment benefit–harm profiles across multiple outcomes and encourages multivariate approaches in applied clinical and policy settings.

## 1. Introduction

Multivariate network meta-analysis (mvNMA) is an extension of network meta-analysis (NMA) that enables the joint synthesis of evidence across multiple treatments and outcomes[1–5]. This capability is particularly relevant for health technology assessment (HTA) and medical decision-making, which often seek to evaluate treatment benefit–harm profiles across multiple outcomes. By jointly combining indirect evidence and correlated outcomes, mvNMA integrates a larger body of evidence that is directly relevant to decision-makers compared with outcome-specific NMA models. Well-documented advantages of mvNMA include increased precision in treatment effect estimation[3,6–8], mitigation of outcome-reporting bias[9–11], and the ability to evaluate surrogate outcomes[12]. Despite its strong theoretical appeal, the use of mvNMA in practice remains limited[13]; which can also be attributed to the lack of user-friendly software. Consequently, independent outcome-specific NMA models that ignore across-outcome correlations are often preferred when assessing treatment benefit–harm profiles.

Interpreting mvNMA outputs across multiple outcomes can be challenging, as they typically comprise numerous treatment effect estimates with varying levels of uncertainty. For example, even in a simple mvNMA setting comparing seven treatments across two outcomes, the full output consists of 42 treatment effect estimates. In this context, summarizing evidence by extending commonly used NMA ranking metrics, such as the probability of being best ($p_{best}$), and the surface under the cumulative ranking curve[14] (SUCRA), to mvNMA only partially simplifies interpretation. Although these outcome-specific metrics effectively address distinct hierarchy questions[15] within each outcome, relying solely on them when interest simultaneously spans several outcomes yields multiple, and often conflicting, treatment hierarchies complicating decision-making and HTA[16,17].

Based on independent outcome-specific NMA models, previous work has proposed several approaches to facilitate the assessment of treatment benefit–harm profiles. Mavridis et al.[16] extended P-scores, the frequentist analogue of SUCRAs, to a multivariate ranking metric, here termed mP-scores, which yields an amalgamated treatment hierarchy across all outcomes of interest. This approach addresses the treatment hierarchy question "Which treatment outperforms the largest proportion of treatments across all outcomes?", thereby focusing on the overall benefit associated with each treatment across outcomes. Daly et al.[17] proposed spie charts to combine outcome-specific treatment hierarchies into a single metric; however, as recommended by the authors, this method should be applied only to outcomes within the same domain (e.g., efficacy outcomes only). This restriction reflects that using spie charts to combine both efficacy and safety outcomes can obscure the suboptimal performance of some treatments within a specific outcome domain[17]. Building on spie charts, Chiocchia et al.[18] introduced the standardized area within a spie chart (SAWIS) to explicitly create trade-offs between efficacy and safety outcomes. Using absolute NMA treatment effect estimates, Tervonen et al.[19] employed a stochastic multicriteria acceptability analysis (SMAA) approach that yields a rank acceptability index for each treatment, thereby not directly producing a treatment hierarchy, but an output resembling a combined across-outcomes rankogram. Finally, Rücker and Schwarzer[20] proposed a partial-ranking method based on a Hasse diagram, which uses the ordering of treatments within each outcome-specific hierarchy to identify treatment orderings that hold across all outcomes, as well as sets of treatments that remain incomparable.

In this article, we propose a novel mvNMA-based framework to enable the assessment of treatment benefit–harm profiles by producing an across-outcomes treatment hierarchy. To this end, we adapt the VIKOR algorithm[21,22] to the mvNMA setting. VIKOR is a distance-based deterministic algorithm that has been widely used for multicriteria decision analysis (MCDA) in fields such as manufacturing and supply chain management[23]. Adapting this algorithm to the mvNMA setting requires treating multiple outcomes as multiple criteria. The algorithm quantifies both the overall and the worst-case performance of each treatment across outcomes. By balancing these two aspects, the overarching goal of VIKOR is to yield a treatment hierarchy that ranks higher those treatments that offer an optimal compromise between their overall and worst-case performance across all outcomes of interest. In addition, by imposing explicit mathematical decision rules, the method allows the identification of either a unique treatment or a set of treatments that offer the optimal compromise between overall and worst-case performance.

In this article, we describe our mvNMA-based framework based on the so-called hybrid mvNMA model which has been previously proposed as an alternative to the fully-hierarchical parametrization[3,10]. The connection between the two models is shown in our Supplementary Material. Briefly, the key difference between the two mvNMA approaches is that the fully-hierarchical model separately considers within- and across-study outcome correlations, requiring within-study correlations to be extracted from individual studies, whereas the hybrid model combines them into a single ‘hybrid’ parameter estimated from the data[3,24,25]. This resolves a major obstacle as outcome correlations are rarely reported at the study level, thereby complicating the use of the fully-hierarchical mvNMA model[3,24,25]. To enable estimation of all treatment effects, even for treatments with missing evidence in some outcomes, we incorporate DuMouchel (DM)–type priors[1,26,27] within the hybrid mvNMA framework. Using posterior samples of the mvNMA treatment effects, we derive outcome-specific treatment hierarchies by extending common NMA ranking metrics, namely SUCRA, $p_{best}$, and median ranks. We then adapt the VIKOR algorithm[21,22] to combine the outcome-specific hierarchies into a single across-outcomes ranking. We illustrate the proposed framework in terms of two published networks. The first compares eleven physical therapy treatments for knee osteoarthritis across two outcomes[28], while the second compares nine classes of antidepressants across five outcomes[29]. Finally, we discuss the advantages and limitations of the proposed framework and outline directions for future research.

## 2. Methods

### 2.1. Multivariate network meta-analysis assuming hybrid correlation parameters

Assume a network of $N$ studies comparing $K$ treatments across $J$ outcomes. Let $K_i$ and $J_i$ denote the total number of treatment arms and outcomes in study $i$, respectively, $i = 1,2, \dots N$. For mathematical convenience, we assume that all $J$ outcomes are reported in each of the $N$ studies, so that $J_i = J, \forall i$. We further simplify notation by assuming the existence of a common study reference treatment group $b$ shared across all studies in the network. Let $y_{ij,k}$ and $s^2_{ij,k}$ denote the respective treatment effect and variance extracted from study $i$ for the comparison between treatment $k = 1,2, \dots, K$ and the reference treatment $b, k \neq b$, in outcome $j$.

A mvNMA model assuming a hybrid correlation parameter can be parametrized as

$$\boldsymbol{y} = \mathbf{X}\boldsymbol{\theta} + \boldsymbol{\epsilon}, \boldsymbol{\epsilon} \sim \mathbf{N}(\mathbf{0}, \boldsymbol{\Sigma}) \quad (1)$$

where $\boldsymbol{y} = \left[y_{ij,k}\right]$ is a vector of length $\sum_{i=1}^{N} J(K_i - 1)$, $\boldsymbol{\theta} = \left[\theta_{j,bk}\right]$ is the $J(K-1)$-dimensional vector of basic parameters across all $J$ outcomes, and $\mathbf{X}$ is the model's design matrix of dimension $\sum_{i=1}^{N} J(K_i - 1) \times J(K-1)$ embedding the consistency equations. Following Salanti et al.,[30] matrix $\mathbf{X}$ describes the basic and functional parameters estimated under the contrast-based model (1). Under the assumption of consistency for each outcome $j$, and for any two treatments $l$ and $k$ $(l, k = 1,2, \dots, K)$ we have that

$$\theta_{j,lk} = \theta_{j,bk} - \theta_{j,bl}, \forall j = 1,2, \dots, J \quad (2)$$

Equation (1) assumes an overall variance-covariance matrix $\mathbf{\Sigma} = \mathbf{diag}[\mathbf{\Sigma}_{\boldsymbol{i}}], i = 1,2, \dots, N$, which is a block-diagonal matrix where each block corresponds to a study $i = 1,2 \dots, N$. Matrices $\mathbf{\Sigma}_{\boldsymbol{i}}, i = 1,2, \dots, N$ have dimension equal to $J(K_i - 1) \times J(K_i - 1)$.

Following previous work[3], we assume that across matrices $\mathbf{\Sigma}_{\boldsymbol{i}}$ all pairwise comparisons share a common outcome-specific heterogeneity parameter $\psi_j, j = 1,2, \dots, J$, to reduce computational burden and estimation difficulties. In addition, we assume a common correlation parameter, namely $\rho_{jz}, j \neq z = 1,2, \dots, J$, across all different pairwise comparisons that share the same pair of outcomes. In the hybrid mvNMA model, these correlation parameters represent an amalgam of within- and across-study outcome correlations[3,24]. In a two-arm study $i$ comparing treatments $k$ and $b$ across two outcomes, the matrix $\mathbf{\Sigma}_{\boldsymbol{i}}$ takes the form

$$\mathbf{\Sigma}_{\boldsymbol{i}} = \begin{pmatrix} \zeta_{i1,k} & \rho_{12}\sqrt{\zeta_{i1,k}\zeta_{i2,k}} \\ \rho_{12}\sqrt{\zeta_{i1,k}\zeta_{i2,k}} & \zeta_{i2,k} \end{pmatrix} \tag{3}$$

where $\zeta_{i1,k} = s_{i1,k}^2 + \psi_1^2$ and $\zeta_{i2,k} = s_{i2,k}^2 + \psi_2^2$. For multi-arm studies, the matrix $\mathbf{\Sigma}_{\boldsymbol{i}}$ is extended to account for two additional types of correlation structures between treatment effects that: (i) evaluate the same outcome and share a common reference group, and (ii) evaluate different outcomes and share a common reference group. More details with respect to the variance-covariance matrix $\mathbf{\Sigma}_{\boldsymbol{i}}$ as well as its extension to accommodate multi-arm studies are available in the Supplementary Material.

#### 2.1.1. Model fit and choice of priors

We adopt a Bayesian framework, assigning priors to the $J(K-1) + J + \binom{J}{2}$ parameters of interest. These are the basic treatment effect parameters $\theta_{j,bk}$, the heterogeneity terms $\psi_j$, and the outcome correlations $\rho_{jz}$. In terms of the treatment effects the standard approach is to assign unrelated non-informative normal priors (e.g. $\theta_{j,bk} \sim \mathrm{N}(0,10^3)$). For a given outcome $j$, this formulation permits estimating treatment effects only between treatments included in at least one study investigating outcome $j$. Although this is assumed in our setting for mathematical convenience, in practice, information on certain treatments may be missing at random across outcomes.

To address the issue of missing treatments across outcomes and estimate all $J(K-1)$ basic treatment effects, a DM prior can replace the non-informative prior for the $\theta_{j,bk}$ parameters. This prior setting assumes that each basic treatment effect parameter $\theta_{j,bk}$ can be decomposed as the sum of a treatment-specific effect $\alpha_k, k = 1,2, \dots, K$ and an outcome-specific effect $\gamma_j, j = 1,2, \dots, J$ such that

$$\theta_{j,bk} \sim \mathrm{N}\left(\alpha_k + \gamma_j, \sigma^2\right) \qquad (4)$$

with $\alpha_k \sim N(0,10^3), \forall k = 1,2, \dots, K$ and $\gamma_j \sim N(0,10^3), \forall j = 1,2, \dots, J$. Equation (4) suggests that each treatment $k = 1,2, \dots, K$ has a constant relative potency across all $J$ outcomes. Under this assumption, the ratio of treatment effects between non-reference treatments remains constant across outcomes on the logarithmic scale. This result follows directly, since, under Equation (2), $\theta_{j,lk} = (\theta_{j,bk} - \theta_{j,bl}) \sim \mathrm{N}(\alpha_k - \alpha_l, 2\sigma^2)$. This implies that treatment effects between non-reference treatments are exchangeable across all outcomes. Consequently, through this exchangeability assumption the DM prior enables information sharing on treatment effects of the same comparison across outcomes, thereby allowing,

a) estimation of missing treatment effects for comparisons between any two treatments, provided that their relative treatment effect is estimable from the data in at least one outcome; and
b) borrowing strength for comparisons with complete data across outcomes, resulting in more precise estimates.

The across-outcomes information sharing implied by Equation (4) requires outcomes to be sufficiently 'similar' and measured on a common scale. Allowing information sharing between efficacy and safety outcomes will often be inappropriate, as these represent distinct outcome domains. Similarly, when continuous outcomes are measured on different scales, information sharing should only be considered after ensuring outcome comparability (e.g., by expressing outcomes on a standardised scale).

To restrict information sharing to the most similar outcomes, we modify Equation (4) accordingly. Let $G$ denote the number of outcome subgroups within which outcomes are sufficiently similar to permit meaningful information sharing. Equation (4) can then be reformulated as, $\theta_{j,bk} \sim \mathrm{N}(\alpha_k^{(g)} + \gamma_j^{(g)}, \sigma_{(g)}^2)$, where $g = 1,2, \dots, G$. This formulation assigns a domain-specific constant relative potency assumption within each outcome subgroup separately.

The standard deviation parameter $\sigma$ in Equation (4) signifies the plausibility of the constant relative potency assumption under the DM prior setting. Small $\hat{\sigma}$ values suggest that the data support this assumption, while larger values indicate the opposite. Following the literature[1], we assign a non-informative prior for this parameter (e.g., $\sigma \sim \mathrm{HN}(0, 10^2)$). Under the DM prior setting, the estimation problem remains identifiable only when no more than $[J(K-1)] - [J + (K-1)]$ treatment effects are missing across all outcomes. This restriction arises because estimation relies on $J(K-1)$ equations to estimate $J$ outcome-specific parameters $\gamma_j$ and $K-1$ treatment-specific parameters $\alpha_k$.

For heterogeneity parameters $\psi_j$, following Röver et al.[31], we assume weakly informative half-normal priors such that $\psi_j \sim \mathrm{HN(c)}, j = 1,2, \dots N, c > 0$. Uniform priors can be assigned to the correlation parameters, such that $\rho_{jz} \sim \mathrm{Unif}(a, b), \forall j, z = 1,2, \dots, J$, where $a, b \in [-1,1]$ and $a < b$.

## 2.2. Producing outcome-specific treatment hierarchies in mvNMA

We derive outcome-specific treatment hierarchies by adapting standard NMA ranking metrics to the mvNMA framework. Specifically, for a beneficial outcome $j$ (i.e. higher treatment effect values are desirable) the ranking distribution for each treatment $k = 1,2, \dots K$, denoted $D_j(k)$, can be expressed as

$$D_j(k) = K - \sum_{\substack{l=1 \\ k \neq l}}^{K} \mathrm{I}_{\{d_{j,bk} - d_{j,bl} > 0\}} \tag{5}$$

where $d_{j,bk}$ are posterior samples for the comparison between treatment $k$ and the reference treatment $b$, and $\mathrm{I}_{\{x\}}$ is an indicator equal to 1 when $d_{j,bk} - d_{j,bl} > 0$ and 0 otherwise. The probability of each treatment $k$ being ranked in position $h = 1,2 \dots, K$ in outcome $j$ is then

$$P_j(k, h) = \Pr\big(D_j(k) = h\big), \forall k, h = 1,2.., K \text{ and } j = 1,2, \dots, J \tag{6}$$

Setting $h = 1$ in Equation (6) yields the $p_{best}$ ranking metric for each treatment $k$ in outcome $j$. Similarly, SUCRA values for each treatment $k$ in outcome $j$ can be calculated from Equation (6) as $\mathrm{SUCRA}(j, k) = \frac{1}{K-1} \sum_{h=1}^{K-1} \sum_{r=1}^{h} P_j(k, r)$. Finally, we estimate the median rank of each treatment $k$ in outcome $j$, denoted as $\mathrm{m_{jk}}$, by taking the median of $D_j(k)$ from Equation (5). Median ranks are non-probabilistic ranking metrics and are typically reported with 95% credible intervals (CrI), for each $\mathrm{m_{jk}}$, thereby reflecting uncertainty in the estimated treatment ranks.

## 2.3. Compromise treatment hierarchies across multiple outcomes using the VIKOR method

While ranking metrics from Section 2.2 summarize treatment performance within each outcome, treatment hierarchies often conflict across outcomes. To address this issue, we adapt the VIKOR method, originally proposed in the field of MCDA[21,22], to the mvNMA setting. The objective of VIKOR in this context is to balance between the overall and worst-case treatment performance and identify the optimal ‘compromise treatment option(s)’ across multiple treatments and outcomes. This can be expressed either as a unique treatment choice or a set of treatments.

The compromise solution can only be identified from treatments included in every outcome-specific hierarchy; treatments absent from at least one hierarchy are excluded from consideration. This issue arises only when the mvNMA model is fitted with a non-informative prior for the treatment effect parameters, since under the DM prior it is always possible to estimate all $J(K-1)$ treatment effects. To produce an across-outcomes treatment hierarchy and derive the set of compromise treatment options, the VIKOR method employs a five-step, distance-based deterministic algorithm. In the remainder of this section, we describe the steps of the adapted VIKOR algorithm for the SUCRA ranking.

***Step 1: Determine the outcome-specific performance measure of interest and the positive ideal solution across all outcomes***

Let $\text{SUCRA}(j,k)$ denote the respective ranking metric of interest summarizing the performance of treatment $k=1,2,\dots,K$ in terms of outcome $j=1,2,\dots,J$. The best positive ideal solution corresponds to a SUCRA value equal to 1 across all outcomes. This value refers to a hypothetical treatment achieving the ideal performance across all outcomes.

***Step 2: Define the ‘overall’ and ‘worst-case’ treatment performance across all J outcomes of interest***

Following Duckstein et al.[32], we obtain the ‘overall’ and ‘worst-case’ treatment performance across all outcomes based on expressions of the $L_p$-norm, defined for each treatment $k=1,2,\dots,K$ as

$$L_{p,k}=\left\{\sum_{j=1}^{J}\left[w_j\left(1-\text{SUCRA}(j,k)\right)\right]^p\right\}^{1/p} \quad (7)$$

where $w_j$ represents the relative weight assigned to outcome $j$, reflecting its importance in the decision-making process, with $\sum_{j=1}^{J} w_j=1$. Assigning values to the weights $w_j$ stems from eliciting clinical or patient preferences. Recently, Mavridis et al.[33], proposed conjoint analysis and a multidimensional scaling approach to elicit and numerically express such outcome weights. In absence of outcome-preferences, we can assume equal outcome weights, such that $w_j = \frac{1}{J}$. Further details on defining outcome-specific weights are provided elsewhere[33] and are beyond the scope of this article. Setting $p = 1$ and $p = \infty$ in Equation (7) yields the overall and worst-case performance of treatment $k$ across all outcomes, denoted as $S_k$ and $R_k$ respectively, defined as

$$S_k = \sum_{j=1}^{J} w_j \big(1 - \text{SUCRA}(j,k)\big) \tag{8}$$

and

$$R_k = \max_j \big[w_j \big(1 - \text{SUCRA}(j,k)\big)\big] \tag{9}$$

where both $S_k$ and $R_k$ range between 0 and 1. Both $S_k$ and $R_k$ represent two aspects of each treatment's proximity to the ideal solution. Equation (8) aggregates the weighted distances between the SUCRA values of treatment $k$ from the positive ideal solution across all outcomes. Thus, $S_k$ represents the overall performance of treatment $k$ across outcomes. In contrast, Equation (9) captures the maximum weighted distance between treatment $k$ and the positive ideal solution across all $J$ outcomes. Consequently, metric $R_k$ describes the worst-case performance associated with treatment $k$ across outcomes. The use of weights in Equations (*8*)-(9) allows determining the overall and worst-case treatment performance based on the importance of each outcome to the decision-maker.

***Step 3: Compromise between overall and worst-case treatment performance***

Following Opricovic et al.[21,22], a compromise between overall and worst-case performance of treatment $k$ across all outcomes, denoted as $Q_k$, is defined as

$$Q_k = v \frac{S_k - \min_k S_k}{\max_k S_k - \min_k S_k} + (1 - v) \frac{R_k - \min_k R_k}{\max_k R_k - \min_k R_k} \tag{10}$$

where $Q_k \in [0,1]$. The compromise metric $Q_k$ integrates both $S_k$ and $R_k$, balanced by the variable $v \in [0,1]$. This variable reflects the 'weight of the decision-making process', as specified by the decision maker. Values of $v$ close to 1 prioritize overall over worst-case performance. Typically, $v$ is set equal to 0.5, to allow for a decision-making strategy that balances both aspects equally.

***Step 4: Rank all treatments according to ranking lists S, R and Q***

Gathering the values $S_k$, $R_k$ and $Q_k, \forall k = 1,2,\dots,K$ yield the ranking lists based on metrics $S$, $R$ and $Q$. Lower values within each ranking list indicate greater proximity to the ideal solution, and thus better treatment performance. Metric $Q$ is typically of primary interest, as it embeds information on overall and worst-case treatment performance, outcome-specific weights, and the decision-making strategy.

***Step 5: Identify the optimal compromise treatment option(s) across all outcomes***

The treatment $k = 1,2,\dots,K$ ranked at the first position by metric $Q$, denoted as $Q(k,1)$, is the unique compromise solution across all $J$ outcomes if it satisfies the conditions of "acceptable advantage" and "acceptable stability"[21,22]:

> **C1. Acceptable advantage:** $Q(h,2) - Q(k,1) \geq \frac{1}{K-1}$ where $Q(h,2)$ is the $Q$-value for treatment $h \neq k = 1,2,\dots,K$ ranked second.
>
> **C2. Acceptable stability:** Treatment $k$ is ranked first by both $S$ and $R$.

The acceptable advantage condition (C1) ensures that the treatment ranked first in the $Q$ metric is sufficiently distinct from the treatment ranked second. The threshold specified in C1 is motivated by the rationale that, if all $Q_k$ values were equally spaced between 0 and 1, the expected difference between consecutive values would be $\frac{1}{K-1}$. The acceptable stability condition (C2) ensures that the identified compromise treatment option is consistent across both aspects of the decision-making process

If one or both of the conditions C1 and C2 are not satisfied, a set of compromise treatment options is proposed:

- the set includes the two treatments $k$ and $h$ ranked first and second by the $Q$ metric, if only C2 is not satisfied;
- the set includes up to $M \leq K$ treatments, if C1 is not satisfied. Treatments are progressively added to the compromise set as long as $Q(h,M) - Q(k,1) < \frac{1}{K-1}, \forall h = 1,2,\dots K$, where $Q(h,M)$ refers to the $Q$ metric value of the treatment $h$ ranked in the $M^{th}$ position.

### 2.4. Extensions of the VIKOR method

Harmonizing the VIKOR method presented in Section 2.3 with other probabilistic outcome-specific ranking metrics such as the $p_{best}$ is straightforward by replacing $\text{SUCRA}(j,k)$ with $P_j(k,1)$ in Equations (7) to (9).

We further deal with outcome-specific ranking metrics presented alongside a 95% CrI, such as median ranks, by adapting the 'fuzzy' VIKOR approach[34] in the mvNMA setting. Similar to VIKOR, the overarching goal of its fuzzy VIKOR extension is to identify the optimal compromise treatment options across all $J$ outcomes using a distance-based algorithm. The fuzzy VIKOR algorithm differs only slightly from the VIKOR algorithm, as presented in Equations (7) to (10). More details on this algorithm are provided in the Supplementary Material.

## 3. Clinical Examples

We illustrate the proposed framework using two published networks. The first compares eleven physical therapy treatments, including placebo, for community-dwelling adults with knee osteoarthritis across two patient-reported continuous outcomes: pain and disability[28]. The second example concerns a network comparing nine antidepressants, including placebo, across five binary outcomes[29]. The primary outcome is early response, defined as at least a 50% reduction in depression scores following treatment. The four secondary outcomes are early remission, adverse events (AE), loss to follow-up, and loss to follow-up due to AE. Early response and remission represent efficacy outcomes, whereas AE, loss to follow-up, and loss to follow-up due to AE represent safety outcomes.

### 3.1. Analysis settings

For each dataset we fit both mvNMA and independent NMA models. Bayesian mvNMA models were fitted using both the 'standard' (mvNMA(standard)) and the 'DuMouchel' (mvNMA(DM)) approaches as presented in Subsection 2.1.1. A non-informative prior $\theta_{j,bk} \sim \text{N}(0, 10^3)$ was assumed for the basic treatment effect estimates in mvNMA(standard) model. Across all analyses we assumed a $\text{HN}(1)$ prior for the heterogeneity parameters. For the correlation parameters, we assigned either a $\text{Unif}(0,1)$ or a $\text{Unif}(-1,0)$ prior between outcomes of the same and opposite outcome domains. Four MCMC chains of 60,000 iterations were run for each Bayesian model, with the first 15,000 iterations discarded as burn-in. We assessed convergence using the Gelman–Rubin statistic[35], with values below 1.05 indicating convergence, and by visually inspecting traceplots.

Outcome-specific treatment hierarchies were calculated in terms of the SUCRA ranking metric while the best compromise solutions across all outcomes obtained using the proposed VIKOR method with equal outcome weights, $w_j = \frac{1}{J}, \forall j = 1,2, \dots, J$, and a balanced decision-making strategy with $v = 0.5$. These values were selected solely to illustrate the use of the method. The fuzzy VIKOR method was used to synthesize treatment hierarchies obtained in terms of median ranks and 95% CrIs.

To facilitate comparison with existing ranking approaches, we additionally fitted independent frequentist NMA models for each outcome. These models enabled computation of mP-scores[16], which provide an amalgamated treatment hierarchy across outcomes based on independent NMA results. Following the original specification[16], mP-scores were computed using cumulative probabilities from a multivariate normal distribution with mean equal to the vector of outcome-specific NMA estimates and a variance–covariance matrix that either (i) assumes zero outcome correlation or (ii) incorporates ad hoc correlation estimates obtained from the mvNMA model. For the independent frequentist NMA models, outcome-specific hierarchies were expressed using P-scores[36]. Given the theoretical equivalence between SUCRA and P-scores, this allows direct comparison of outcome-specific hierarchies derived from mvNMA and NMA models, with differences attributable primarily to model specification rather than to the ranking metrics per se.

We did not include alternative approaches, such as Hasse diagrams[20] or SMAA[19], as their outputs are not directly comparable to the proposed ranking framework. In addition, because both efficacy and safety outcomes are considered simultaneously, we did not apply the spie chart approach[17]. However, in the Supplementary Material, we reanalyze the psoriasis example[37] presented by Daly et al.[17] to illustrate the robustness of VIKOR compared to spie charts when severe conflicts across outcomes exist. Given that we did not have access to the full dataset, in this case we illustrate the VIKOR approach based only on the SUCRA values reported in the original publication, which were derived from outcome-specific NMA models.

All analyses were conducted in R version 4.6.0 (2026-04-24). To facilitate implementation of the proposed framework, we developed the R package **mvnma**, which automates mvNMA model fitting, estimation of outcome-specific hierarchies, and implementation of the VIKOR algorithm. Frequentist NMA models were fitted using the R package **netmeta**[38], and code for computing mP-scores was obtained from the Supplementary Material of the original publication[16].

## 3.2. Results

### 3.2.1. Network of physical therapy treatments

This dataset comprises of 54 studies that compare eleven physical therapy treatments, including "Placebo" and "No treatment". The network graphs for the two outcomes of interest are shown in **Figure 1**. The outcome of pain is reported in 51 studies **(Figure 1)** while the evidence around disability (**Figure 1)** is sparser as this outcome is reported in 26 studies. Evidence for all eleven treatments was available for both outcomes. A strong positive outcome correlation of 0.76 [0.57, 0.88] was estimated under both mvNMA models. Heterogeneity estimates were also similar across mvNMA approaches, with values of 0.96 [0.74, 1.24] for pain and 0.86 [0.66, 1.12] for disability. Corresponding heterogeneity estimates from independent NMA models were 0.74 and 1.06, respectively.

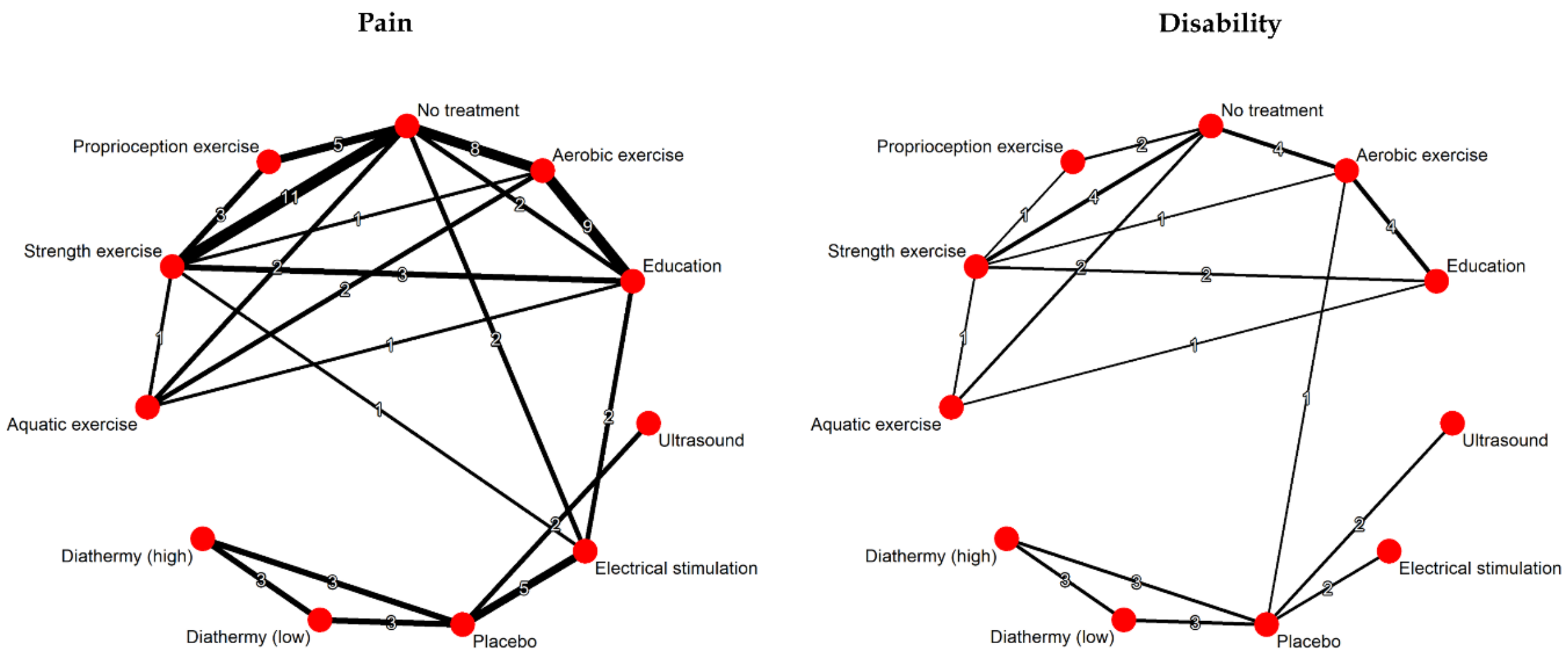

**Figure 1:** Network plot for the two outcomes of interest in the physical therapy treatments network.

Summary mean difference (MD) estimates relative to "No treatment" are presented in **Figure 2**. For pain, mvNMA(standard) and independent NMA models produced very similar point estimates and uncertainty (**Figure 2**), while some differences in the magnitude of treatment effects were observed relative to the mvNMA(DM) approach. For disability, both mvNMA models yielded more precise estimates than independent NMA (**Figure 2**), while maintaining consistent effect directions. The greatest precision was observed under the mvNMA(DM) model, as a consequence of the additional exchangeability assumption. The parameter $\sigma$ was estimated as 0.19 [0.01, 0.49], supporting the constant relative potency assumption. Convergence diagnostics are reported in Supplementary Table 1 and Supplementary Figures 1-

2. Both the R-hat statistics and the visual inspection of traceplots indicate satisfactory convergence for all parameters.

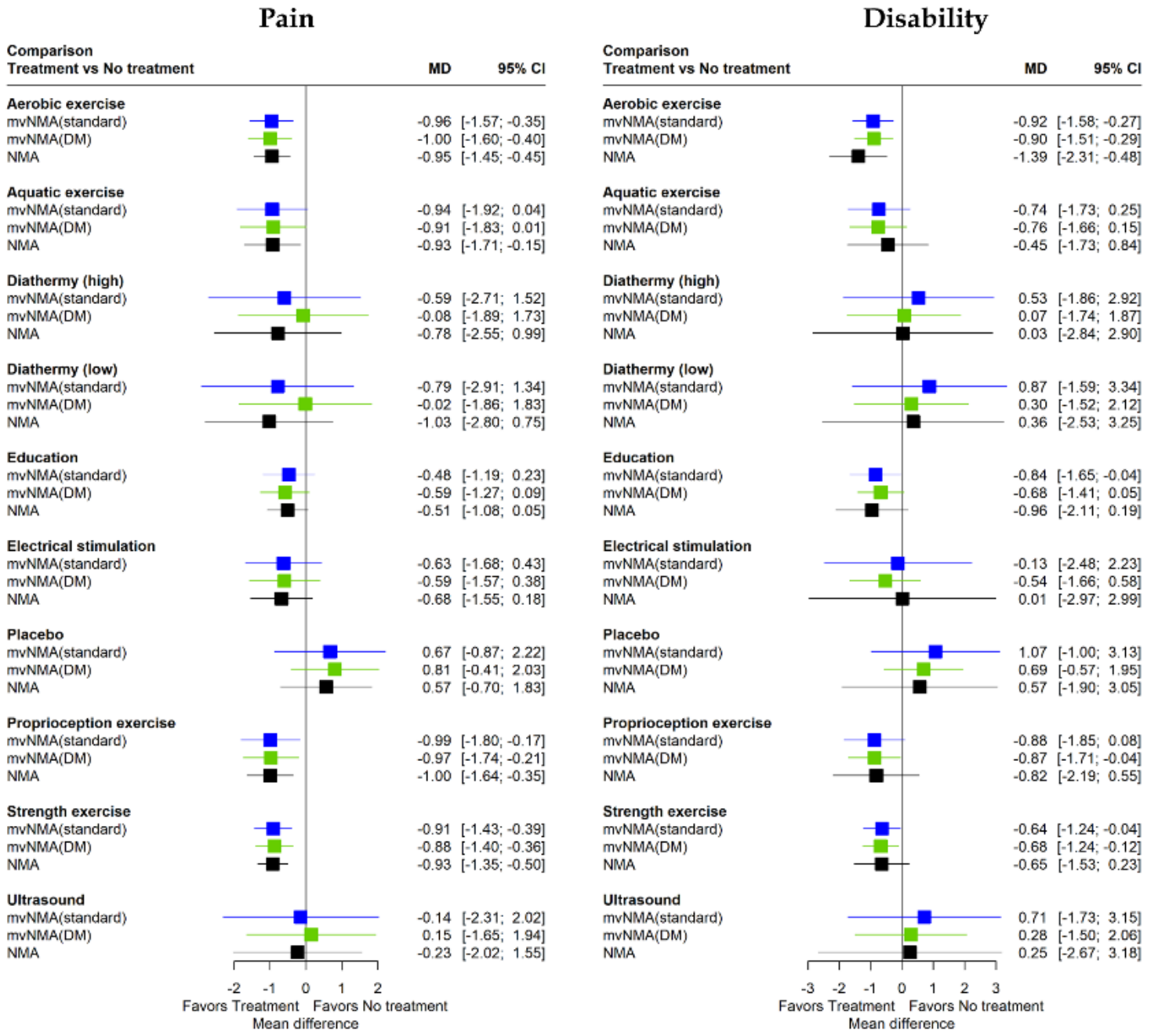

**Figure 2:** Summary treatment effect estimates in terms of the two outcomes of interest for mvNMA(standard), mvNMA(DM), and independent outcome-specific NMA models in the physical therapy treatments network.

Outcome-specific treatment hierarchies are shown in **Table 1**. Treatment hierarchies for pain were consistent across models, whereas some disagreement between mvNMA and outcome-specific NMA models was observed for disability. For example, Proprioception Exercise ranked second under mvNMA but third under independent NMA. In addition, the percentage of treatments ranked worse than Aquatic Exercise was 52% based on independent NMA, increasing to 69% and 70% under the mvNMA(standard) and mvNMA(DM) models, respectively. These differences can be attributed to changes in both the magnitude and uncertainty of the summary treatment effect estimates between the mvNMA and NMA models, as shown in **Figure 2**. This finding highlights that accounting for outcome correlation through the mvNMA modelling framework may lead to different conclusions compared with fitting independent NMA models for each outcome, which implicitly ignores their potential correlation.

**Table 1:** Outcome-specific treatment hierarchies from mvNMA and NMA models for the physical therapy treatments network. Treatments ranking in the top three for each metric are shown in bold. The "Treatment" column is ordered by mvNMA(standard) SUCRA values for the Pain outcome.

| Treatment | Pain | | | Disability | | |
|---|---|---|---|---|---|---|
| | SUCRA mvNMA(standard) | SUCRA mvNMA(DM) | P-score (NMA) | SUCRA mvNMA(standard) | SUCRA mvNMA(DM) | P-score (NMA) |
| Aerobic exercise | **0.72** | **0.80** | **0.70** | **0.80** | **0.80** | **0.88** |
| Proprioception exercise | **0.71** | **0.77** | **0.71** | **0.76** | **0.77** | **0.65** |
| Strength exercise | **0.68** | 0.72 | 0.68 | 0.65 | 0.66 | 0.60 |
| Aquatic exercise | 0.68 | **0.73** | 0.67 | **0.69** | **0.70** | 0.52 |
| Diathermy (low) | 0.60 | 0.36 | **0.68** | 0.22 | 0.27 | 0.33 |
| Electrical stimulation | 0.53 | 0.57 | 0.53 | 0.54 | 0.60 | 0.44 |
| Diathermy (high) | 0.53 | 0.39 | 0.57 | 0.34 | 0.37 | 0.44 |
| Education | 0.42 | 0.54 | 0.40 | 0.75 | 0.66 | **0.71** |
| Ultrasound | 0.37 | 0.30 | 0.35 | 0.27 | 0.28 | 0.37 |
| No treatment | 0.20 | 0.26 | 0.16 | 0.36 | 0.29 | 0.33 |
| Placebo | 0.06 | 0.05 | 0.04 | 0.13 | 0.09 | 0.23 |

Across-outcomes hierarchies derived using VIKOR are presented in **Table 2**. Under both mvNMA approaches, Aerobic Exercise, Proprioception Exercise, and Aquatic Exercise ranked highest according to the $Q$ metric. Based on the VIKOR criterion, these three treatments also constitute the optimal set of compromise treatment options. Although Aerobic Exercise ranks first in terms of both the outcome-specific treatment hierarchies and the amalgamated hierarchy indicated by the $Q$ metric, it does not fulfil the acceptable advantage condition (C1) against either Proprioception Exercise or Aquatic Exercise (i.e. $0.02 - 0.00 < \frac{1}{10}$, and, $0.08 - 0.00 < \frac{1}{10}$ ). Results from the fuzzy-VIKOR analysis and median-rank hierarchies are provided in Supplementary Tables 2–4. The mP-scores approach ranked Aerobic Exercise first, both when ignoring outcome correlation and when incorporating the mvNMA-derived correlation estimate ($\hat{\rho}_{jz} = 0.76$). Incorporating correlation increased the mP-score of Proprioception Exercise, narrowing the difference between the two treatments.

**Table 2:** Treatment rankings across the pain and disability outcomes in the physical therapy treatments network. Rankings from the mvNMA(standard) and mvNMA(DM) approaches were obtained using the VIKOR method, based on ranking lists Q, S and R , whereas rankings from independent NMA models were derived using the mP-scores approach. For the VIKOR method, lower values of Q, S and R indicate better treatment performance. Treatments identified as the optimal compromise solution by the VIKOR algorithm applied to the mvNMA(standard) output are shown in bold. The "Treatment" column is ordered by Q metric values obtained from the mvNMA(standard) approach.

| Treatment | mvNMA(standard) | | | mvNMA(DM) | | | NMA ($\hat{\rho}_{jz} = 0$) | NMA ($\hat{\rho}_{jz}$: mvNMA) |
|---|---|---|---|---|---|---|---|---|
| | $Q$ | $S$ | $R$ | $Q$ | $S$ | $R$ | mP-scores | mP-scores |
| **Aerobic exercise** | **0.00** | 0.24 | 0.14 | 0.00 | 0.20 | 0.10 | **0.62** | **0.67** |
| **Proprioception exercise** | **0.02** | 0.27 | 0.14 | 0.05 | 0.23 | 0.12 | **0.48** | **0.56** |
| **Aquatic exercise** | **0.08** | 0.31 | 0.16 | 0.13 | 0.28 | 0.15 | **0.37** | **0.45** |
| Strength exercise | 0.12 | 0.34 | 0.18 | 0.17 | 0.31 | 0.17 | 0.43 | 0.51 |
| Electrical stimulation | 0.31 | 0.46 | 0.23 | 0.30 | 0.42 | 0.21 | 0.26 | 0.35 |
| Education | 0.35 | 0.41 | 0.29 | 0.31 | 0.40 | 0.23 | 0.33 | 0.38 |
| Diathermy (high) | 0.53 | 0.57 | 0.33 | 0.58 | 0.62 | 0.31 | 0.27 | 0.37 |
| Diathermy (low) | 0.64 | 0.59 | 0.39 | 0.69 | 0.68 | 0.36 | 0.24 | 0.32 |
| Ultrasound | 0.67 | 0.68 | 0.36 | 0.70 | 0.71 | 0.36 | 0.15 | 0.24 |
| No treatment | 0.75 | 0.72 | 0.40 | 0.72 | 0.72 | 0.37 | 0.10 | 0.14 |
| Placebo | 1.00 | 0.91 | 0.47 | 1.00 | 0.93 | 0.47 | 0.01 | 0.04 |

### 3.2.2. Network of antidepressants

This dataset includes 59 studies comparing nine antidepressant classes, including placebo. Network graphs for all outcomes are shown in **Figure 3**. Networks for early response, early remission, loss to follow-up, and loss to follow-up due to AE included all nine classes, whereas the AE network was sparser, comprising 38 studies and excluding class NRI. Outcome correlation estimates from both mvNMA approaches are presented in **Table 3** and indicate low to moderate correlation. Outcome-specific heterogeneity estimates were similar across mvNMA and independent NMA models and are reported in Supplementary Table 5.

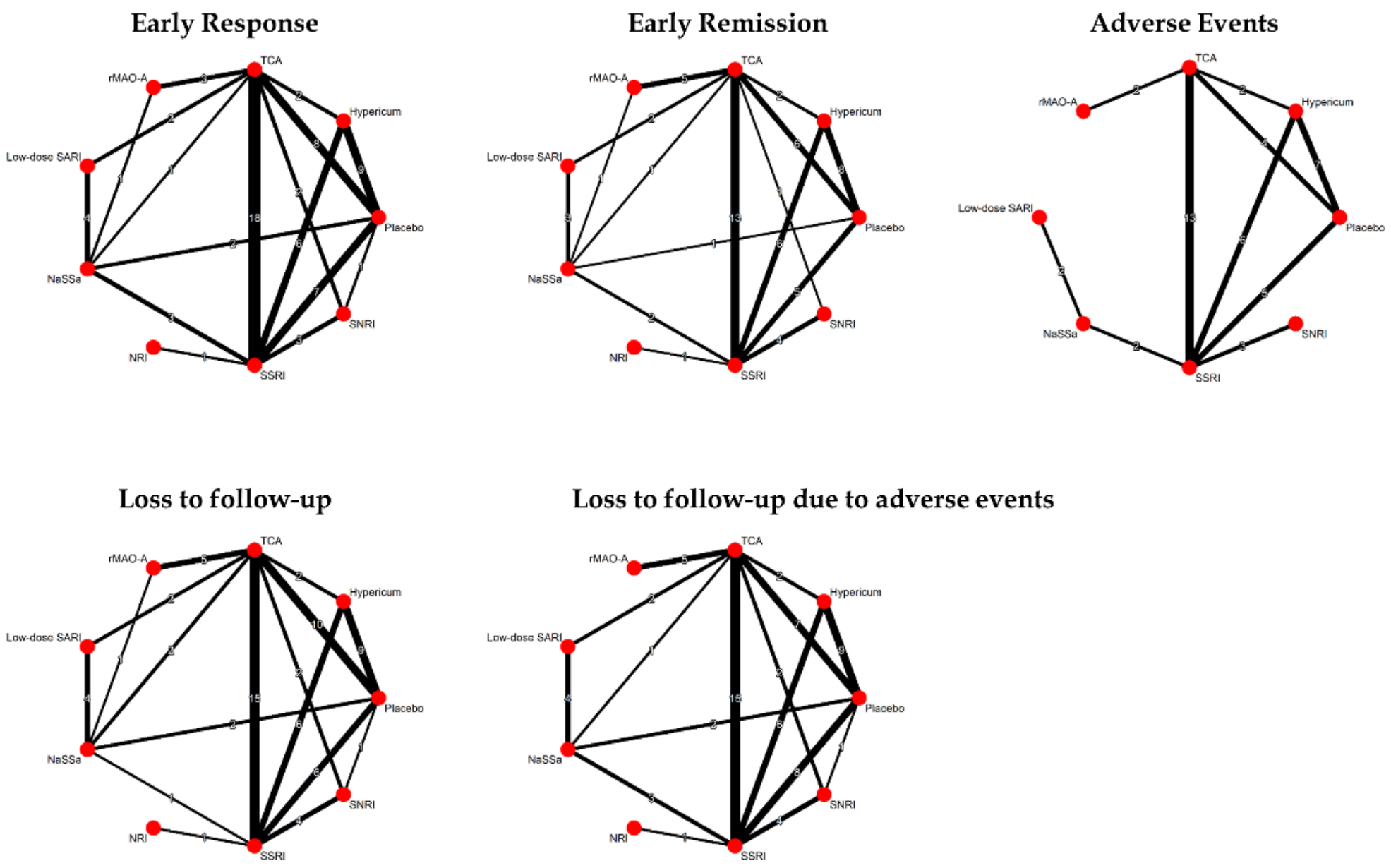

**Figure 3:** Network plot across all five outcomes of interest in the antidepressants dataset.Abbreviations: NaSSa: Noradrenergic and Specific Serotonergic Antidepressants, NRI: Norepinephrine Reuptake Inhibitors, SARI: Serotonin Antagonist and Reuptake Inhibitors, TCA: Tricyclic Antidepressants, SSRI: Selective Serotonin Reuptake Inhibitors, SNRI: Serotonin–Norepinephrine Reuptake Inhibitors, rMAO-A: Reversible Monoamine Oxidase A Inhibitors.

Summary odds ratio estimates relative to placebo are shown in **Figure 4**. The mvNMA(DM) model yielded the most precise estimates, with a median standard error of 0.22. As in this example there are both efficacy and safety outcomes, the mvNMA(DM) model was adjusted to allow for separate information sharing within each outcome domains, restricting information sharing across efficacy and safety. The parameters $\sigma_{(1)}$ and $\sigma_{(2)}$ were estimated as 0.04 [0.00, 0.13] and 0.15 [0.04, 0.31], respectively, thus supporting the constant relative potency assumption within each outcome domain. mvNMA(standard) and independent NMA models produced similar precision, with median standard errors of 0.25 and 0.24, respectively. Notably, summary ORs for treatment NRI with respect to the AE outcome were estimable only under the mvNMA(DM) approach (**Figure 4**), highlighting the ability of this model to borrow information across outcomes with complete data on class NRI and to extrapolate this information to estimate the respective missing treatment effects for the AE outcome. Convergence diagnostics are reported in Supplementary Tables 6–7 and Supplementary Figures 3-4. Overall, based on both R-hat statistics and the visual inspection of traceplots we conclude that MCMC convergence was achieved across all parameters of interest.

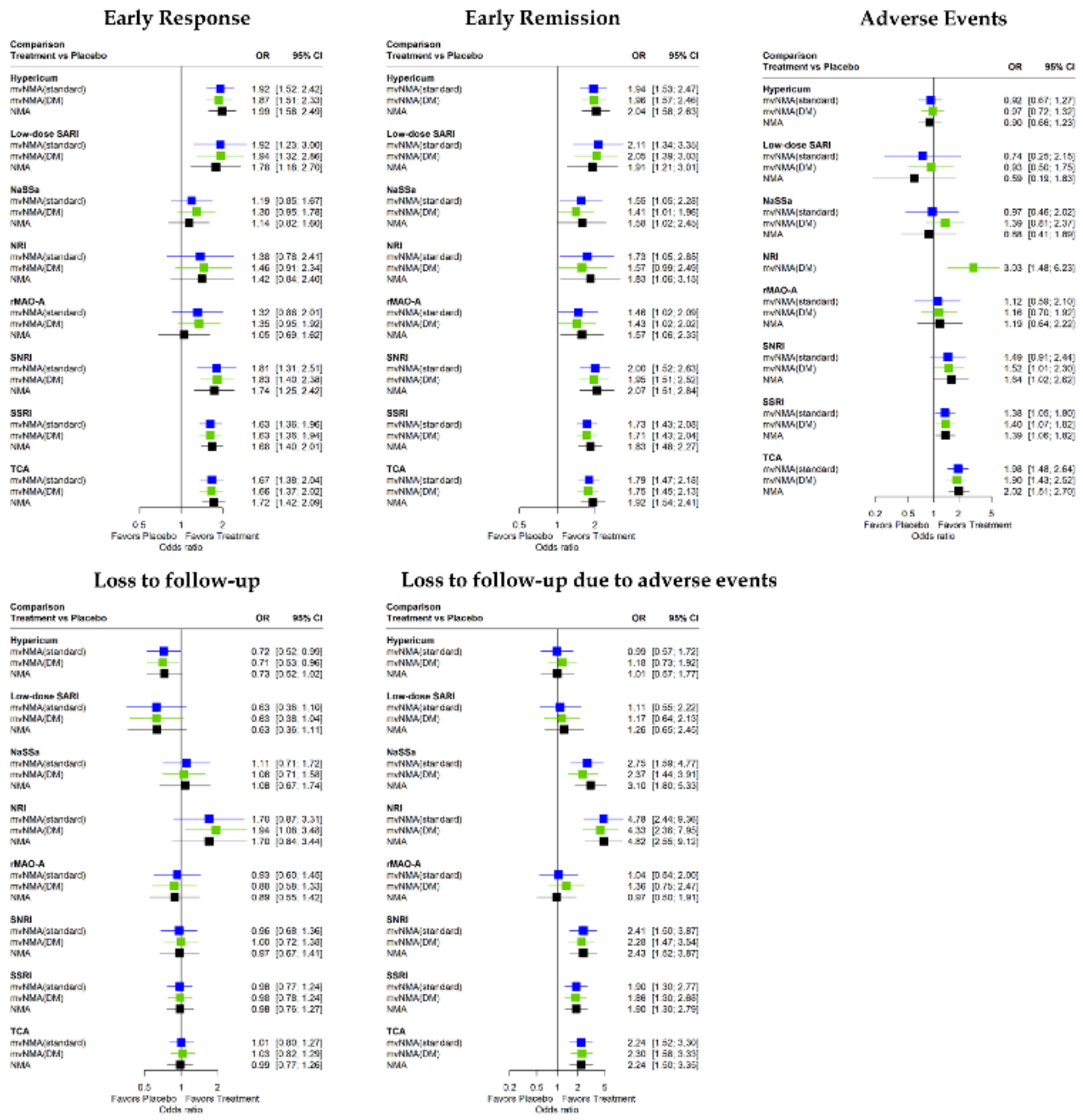

**Figure 4:** Summary treatment effect estimates in terms of all five outcomes of interest for mvNMA(standard), mvNMA(DM), and independent outcome-specific NMA models for the antidepressants dataset. Abbreviations: NaSSa: Noradrenergic and Specific Serotonergic Antidepressants, NRI: Norepinephrine Reuptake Inhibitors, SARI: Serotonin Antagonist and Reuptake Inhibitors, TCA: Tricyclic Antidepressants, SSRI: Selective Serotonin Reuptake Inhibitors, SNRI: Serotonin–Norepinephrine Reuptake Inhibitors, rMAO-A: Reversible Monoamine Oxidase A Inhibitors.

Outcome-specific treatment hierarchies are presented in **Table 4**. Rankings derived across all models were largely consistent, with only minor differences observed. Low-dose SARI and Hypericum exhibited stable, favourable performance across all outcomes, whereas the remaining treatments displayed more conflicted benefit–harm profiles. For example, although class SNRI performed well with respect to the efficacy outcomes of early response and early remission, it performed poorly across all three safety outcomes, ranking among the lowest positions.

Across-outcomes hierarchies derived using VIKOR are shown in **Table 5**. Under both mvNMA approaches, the VIKOR method produced the same amalgamated hierarchy based on the compromise ranking metric $Q$, ranking Low-dose SARI first. Under the mvNMA(DM) approach, the amalgamated treatment hierarchy includes all available treatment classes, including NRI, again reflecting the model's ability to estimate treatment effects for treatments with missing evidence in certain outcomes. Evaluation of the VIKOR conditions C1 and C2 indicates that no unique compromise treatment exists under either mvNMA model. Instead, the optimal compromise set of treatments is shared by Low-dose SARI and Hypericum. Low-dose SARI ranks first according to the $Q$, $S$, and $R$ metrics and thus satisfies the acceptable stability condition (C2); however, it does not meet the acceptable advantage condition (C1), leading to the inclusion of Hypericum in the final compromise set. This result reflects VIKOR's emphasis on balancing overall and worst-case performance across outcomes: treatments with stable outcome-specific performance are favoured over those excelling in only one outcome domain, such as SNRI or rMAO-A. Results based on median ranks and the fuzzy-VIKOR approach are reported in Supplementary Tables 8–11.

**Table 3:** Correlation estimates and 95% credible intervals for all 10 combinations of the five outcomes of interest in the antidepressants example.

| Outcome 1 vs Outcome 2 | Correlation Estimates [95% CrI] mvNMA(standard) | Correlation Estimates [95% CrI] mvNMA(DM) |
|---|---|---|
| Early response vs Early remission | 0.59 [0.38, 0.73] | 0.61 [0.43, 0.75] |
| Early response vs Adverse Events | -0.23 [-0.44, -0.03] | -0.24 [-0.43, -0.04] |
| Early response vs Loss to follow up | -0.40 [-0.58, -0.18] | -0.40 [-0.57, -0.20] |
| Early response vs Loss to follow up due to adverse events | -0.34 [-0.54, -0.12] | -0.33 [-0.52, -0.11] |
| Early remission vs Adverse Events | -0.08 [-0.26, 0.00] | -0.09 [-0.29, 0.00] |
| Early remission vs Loss to follow up | -0.40 [-0.58, -0.19] | -0.40 [-0.58, -0.18] |
| Early remission vs Loss to follow up due to adverse events | -0.30 [-0.51, -0.07] | -0.29 [-0.49, -0.06] |
| Adverse Events vs Loss to follow up | 0.22 [0.02, 0.46] | 0.22 [0.02, 0.45] |
| Adverse events vs Loss to follow up due to adverse events | 0.44 [0.18, 0.64] | 0.44 [0.21, 0.63] |
| Loss to follow up vs Loss to follow up due to adverse events | 0.55 [0.36, 0.69] | 0.54 [0.35, 0.68] |

Both mP-scores approaches yielded an amalgamated treatment hierarchy in which Hypericum and Low-dose SARI ranked at the top (**Table 5**). Interestingly, mP-scores values for most of the remaining classes were close to zero, and even for Hypericum the corresponding mP-scores values of 0.37 and 0.41 are relatively low to justify its first position in the amalgamated hierarchy. Interpretation of these values indicates that, on average, only 37% and 41% of treatments classes performed worse than Hypericum, implying that the majority of classes were not outperformed by Hypericum across all outcomes. This finding reflects the challenges inherent in the conceptual objective of the mP-scores approach to produce an amalgamated treatment hierarchy in the presence of negatively correlated outcomes while quantifying only the overall performance of each treatment.

**Table 4:** Outcome-specific treatment hierarchies from mvNMA and NMA models for the antidepressants example. Treatments ranking in the top three for each metric are shown in bold. The "Treatment" column is ordered by mvNMA(standard) SUCRA values for the Early Response outcome.

| Treatment | Early Response | | | Early Remission | | |
|---|---|---|---|---|---|---|
| | SUCRA mvNMA(standard) | SUCRA mvNMA(DM) | P-score | SUCRA mvNMA(standard) | SUCRA mvNMA(DM) | P-score |
| Hypericum | **0.84** | **0.82** | **0.89** | **0.73** | **0.81** | **0.75** |
| Low-dose SARI | **0.80** | **0.82** | **0.72** | **0.80** | **0.82** | 0.62 |
| SNRI | **0.75** | **0.78** | **0.69** | **0.79** | **0.80** | **0.77** |
| TCA | 0.62 | 0.60 | 0.68 | 0.57 | 0.59 | **0.65** |
| SSRI | 0.55 | 0.53 | 0.62 | 0.47 | 0.51 | 0.51 |
| NRI | 0.38 | 0.41 | 0.44 | 0.52 | 0.43 | 0.55 |
| rMAO-A | 0.31 | 0.28 | 0.15 | 0.27 | 0.28 | 0.32 |
| NaSSa | 0.20 | 0.24 | 0.21 | 0.35 | 0.26 | 0.33 |
| Placebo | 0.05 | 0.02 | 0.09 | 0.01 | 0.01 | 0.01 |
| | **Adverse Events** | | | **Loss to follow up** | | |
| Hypericum | **0.75** | **0.83** | **0.75** | **0.87** | **0.88** | **0.85** |
| Low-dose SARI | **0.81** | **0.83** | **0.88** | **0.92** | **0.93** | **0.91** |
| SNRI | 0.26 | 0.35 | 0.19 | 0.51 | 0.45 | 0.47 |
| TCA | 0.03 | 0.14 | 0.03 | 0.41 | 0.36 | 0.44 |
| SSRI | 0.31 | 0.44 | 0.32 | 0.48 | 0.47 | 0.45 |
| NRI | - | 0.02 | - | 0.05 | 0.01 | 0.06 |
| rMAO-A | 0.54 | 0.64 | 0.48 | **0.54** | **0.62** | **0.60** |
| NaSSa | 0.65 | 0.45 | **0.70** | 0.30 | 0.35 | 0.33 |
| Placebo | **0.66** | **0.80** | 0.64 | 0.43 | 0.43 | 0.41 |
| | **Loss to follow up due to adverse events** | | | | | |
| Hypericum | **0.84** | **0.80** | **0.83** | | | |
| Low-dose SARI | **0.77** | **0.80** | **0.71** | | | |
| SNRI | 0.26 | 0.28 | 0.28 | | | |
| TCA | 0.32 | 0.26 | 0.33 | | | |
| SSRI | 0.48 | 0.49 | 0.50 | | | |
| NRI | 0.01 | 0.01 | 0.01 | | | |
| rMAO-A | 0.8 | 0.69 | **0.85** | | | |
| NaSSa | 0.19 | 0.24 | 0.15 | | | |
| Placebo | **0.83** | **0.91** | 0.84 | | | |

## 4. Discussion

In this article, we propose a novel mvNMA-based framework that integrates treatment benefit–harm profiles to identify optimal treatment options across multiple outcomes by adapting the VIKOR MCDA algorithm to the meta-analytic setting. We build our framework based on summary treatment effect estimates and outcome-specific treatment hierarchies obtained from mvNMA, thereby accounting for outcome correlation[1,3,6,7]. This feature distinguishes our approach from much of the existing literature, which, largely for practical reasons, has relied on ranking methods applied to independent, outcome-specific NMAs and has therefore ignored the impact of outcome correlation.

**Table 5:** Treatment rankings across all five outcomes in the antidepressants example. Rankings from the mvNMA(standard) and mvNMA(DM) approaches were obtained using the VIKOR method, based on ranking lists Q, S and R, whereas rankings from independent NMA models were derived using the mP-scores approach. For the VIKOR method, lower values of Q, S and R indicate better treatment performance. Treatments identified as the optimal compromise solution by the VIKOR algorithm applied to the mvNMA(standard) output are shown in bold. The "Treatment" column is ordered by Q metric values obtained from the mvNMA(standard) approach.

| **Treatment** | **mvNMA (standard)** | | | **mvNMA(DM)** | | | **NMA ($\hat{\rho}_{jz} = 0$)** | **NMA ($\hat{\rho}_{jz}$: mvNMA)** |
|---|---|---|---|---|---|---|---|---|
| | Q | S | R | Q | S | R | mP-scores | mP-scores |
| **Low-dose SARI** | **0.00** | 0.19 | 0.05 | 0.00 | 0.16 | 0.04 | **0.24** | **0.27** |
| **Hypericum** | **0.01** | 0.20 | 0.05 | 0.01 | 0.17 | 0.04 | **0.37** | **0.41** |
| rMAO-A | 0.65 | 0.53 | 0.15 | 0.58 | 0.50 | 0.14 | 0.01 | 0.02 |
| SSRI | 0.66 | 0.58 | 0.14 | 0.50 | 0.51 | 0.11 | 0.02 | 0.02 |
| SNRI | 0.72 | 0.52 | 0.17 | 0.57 | 0.47 | 0.15 | 0.01 | 0.02 |
| Placebo | 0.93 | 0.63 | 0.20 | 0.81 | 0.57 | 0.20 | 0.00 | 0.00 |
| TCA | 0.93 | 0.65 | 0.19 | 0.76 | 0.61 | 0.17 | 0.00 | 0.00 |
| NaSSa | 0.95 | 0.70 | 0.18 | 0.76 | 0.69 | 0.15 | 0.00 | 0.00 |
| NRI | - | - | - | 1.00 | 0.82 | 0.20 | - | - |

We present mvNMA using a hybrid correlation parametrization[3,24,25], which simplifies model implementation by requiring the same amount of data as a standard NMA. In addition to the standard mvNMA specification with non-informative normal priors for treatment effects, we adapt DM priors[26] within the hybrid mvNMA model. The DM prior assumes constant relative treatment potency across outcomes, implying exchangeability of treatment effects among non-reference treatments. This assumption enables borrowing of information across treatments and outcomes and allows estimation of all treatment effects, even when evidence for specific treatments is missing for some outcomes. The added benefits of the mvNMA(DM) in terms of estimation and precision come at the cost that the use of the model cannot be as broad as that of the mvNMA(standard) model. Specifically, the borrowing of strength property should only be used after carefully ensuring that the sharing of information across outcomes is meaningful. In addition, a meaningful estimation of missing outcome-specific treatment effect parameters is possible only when no more than $[J(K-1)] - [J + (K-1)]$ treatment effects are missing across all outcomes.

We generalize the most commonly[39] used outcome-specific ranking metrics, namely SUCRA, $p_{BV}$ and median ranks within the mvNMA framework. Each metric addresses a distinct treatment hierarchy question and summarizes treatment performance within a given outcome[15]. However, when multiple outcomes are considered, outcome-specific hierarchies often conflict, complicating decision-making[16,17]. To address this, we adapt the VIKOR MCDA algorithm to the mvNMA setting. VIKOR has previously been applied to synthesize rankings across multiple criteria in fields such as manufacturing and supply chain management[23]. In the mvNMA context, VIKOR functions as a distance-based deterministic algorithm[21,22] that synthesizes outcome-specific treatment hierarchies and balances between overall and worst-case treatment performance across outcomes. The method accommodates hierarchies derived from probabilistic ranking metrics, and, through its fuzzy-VIKOR extension[34], also supports ranking metrics presented alongside uncertainty measures, such as median ranks, which are typically reported alongside 95% CrIs. As a result, VIKOR is compatible with a broad range of outcome-specific hierarchy questions[15]. In addition, Step 1 of the VIKOR method can be extended to incorporate additional treatment-specific criteria beyond outcome-specific performance measures, such as treatment cost[40]. In this case, VIKOR can yield an amalgamated treatment hierarchy while accounting for these additional characteristics.

The primary objective of VIKOR is to yield an across-outcomes treatment hierarchy and to explicitly identify treatments that achieve the optimal compromise between overall and worst-case performance. The method is based on weighted distances defined using the $L_p$-norm. Specifically, the $L_1$ and $L_\infty$ norms yield the $S$ and $R$ metrics, which quantify overall and worst-case treatment performance across outcomes, respectively. VIKOR then defines the compromise metric $Q$, which balances these two aspects. The $S$, $R$ and $Q$ metrics represent different measures of proximity to the ideal solution, with lower values indicating better performance. In the final step, VIKOR evaluates the conditions of acceptable advantage (C1) and acceptable stability (C2) to identify either a unique compromise treatment or a set of treatments with similar balance between their overall and worst-case performance.

Equations (5)-(6) provide the basis for calculating outcome-specific treatment hierarchies based on marginal probabilities that indicate how likely it is that a treatment $k = 1,2, \ldots K$ will rank in a given position $h = 1,2, \ldots, K$ for a given outcome $j = 1,2, \ldots J$. When the interest spans multiple outcomes, these equations can easily be modified to enable the calculation of joint probabilities, thereby extending the complex treatment hierarchy framework originally proposed by Papakonstantinou et al.[41], in the context of mvNMA. Given that in this manuscript our interest mostly lies in producing an across-outcomes treatment hierarchy rather than in answering very specific treatment hierarchy questions such as "what is the probability that treatment ranks first for outcome 1 and first or second for outcome 2," we retained the expressions of Equations (5)-(6) simple, referring only to the case of marginal outcome-specific probabilities.

The literature on MCDA methods for ranking alternatives across multiple outcomes is extensive and includes several approaches[42]. Among these, the method most closely aligned with VIKOR is the Technique for Order Preference by Similarity to Ideal Solution (TOPSIS)[43,44]. TOPSIS requires specification of positive and negative ideal solutions based on outcome-specific performance measures and aggregates information using the Euclidean norm to quantify distances from these ideal points. A relative closeness metric is then computed to integrate distances from both the positive and negative ideal solutions. While TOPSIS is conceptually similar to VIKOR, it produces only an amalgamated ranking, analogous to the $Q$ metric, without explicitly identifying a final set of compromise treatment options. In addition, TOPSIS assigns equal importance to distances from the positive and negative ideal solutions, whereas VIKOR allows decision-makers to control their relative importance through the parameter $v$.

We used two published datasets to illustrate the proposed framework and to compare it with other alternatives. In this manuscript, we use these datasets solely for illustration and methodological comparison, without aiming to directly inform clinical decisions. In practice, the framework should be applied with caution, in collaboration with a multidisciplinary team that first evaluates which treatments and studies to include in the analysis, as well as other potential sources of bias[45] (e.g., publication bias) that should be addressed before producing any ranking.

In the network of physical therapy treatments[28], producing an across outcomes treatment hierarchy proved relatively straightforward, as the analysis involved only two positively correlated outcomes, thereby limiting conflicts across outcome-specific treatment hierarchies. Nevertheless, this example highlights the importance of relying on mvNMA for decision-making in the presence of correlated outcomes and substantial imbalances in the number of studies informing each outcome. Specifically, both mvNMA models yielded more precise estimates for the disability outcome. These differences in the treatment effect estimates for disability translated into differences in the outcome-specific treatment hierarchies, with Aquatic Exercise showing improved performance, yielding larger SUCRA values under the mvNMA models compared with the outcome-specific NMA approach. Using independent outcome-specific NMA models, the amalgamated treatment hierarchy based on mP-scores ranked Aerobic Exercise first. The VIKOR method based on the mvNMA framework produced a similar amalgamated treatment hierarchy in terms of treatment order, but identified a set of compromise treatment options across the two outcomes, comprising Aerobic Exercise, Proprioception Exercise, and Aquatic Exercise, rather than a unique solution.

In the antidepressants dataset[29], the mvNMA(standard) and independent NMA models produced very similar treatment effect estimates and precision, and consequently yielded nearly identical outcome-specific treatment hierarchies. Under the mvNMA(DM) framework, the exchangeability assumption led to more precise treatment effect estimates. In addition, this framework enabled estimation of all treatment effects, including those for treatment NRI, for which evidence was originally missing for the AE outcome. Summarising the extensive output of each method into outcome-specific treatment hierarchies resulted in multiple, and in some cases conflicting, across-outcomes hierarchies under both the mvNMA and independent NMA frameworks. We applied the proposed VIKOR method based on the outcome-specific treatment hierarchies of both the mvNMA(standard) and mvNMA(DM) models for the antidepressants dataset. This produced an across-outcomes treatment hierarchy that ranked Low-dose SARI and Hypericum in the first and second position, respectively, and identified both as optimal compromise treatment options. The mP-scores approach based on independent NMA models yielded an amalgamated hierarchy that placed Hypericum in the first position and Low-dose SARI in the second. The interpretation of the mP-scores values indicates that the majority of treatments were not outperformed by Hypericum across all outcomes, despite its first-place ranking. This underscores the conceptual challenges of constructing treatment hierarchies across negatively correlated outcomes when relying solely on overall performance metrics. In this sense, the mP-scores approach is conceptually equivalent to the VIKOR method when the latter is restricted only to the $S$ ranking metric.

We see several advantages of our proposed framework for identifying optimal treatment options across multiple outcomes. First, the framework builds on mvNMA and therefore explicitly accounts for outcome correlation, while also mitigating the impact of outcome reporting bias. These advantages become more pronounced under the mvNMA(DM) specification, which enables information sharing across outcomes and yields more precise treatment effect estimates for all possible treatment comparisons. This feature addresses a common challenge in constructing across-outcomes treatment hierarchies, namely the presence of missing evidence for specific treatments in some outcomes, which traditionally complicates the assessment of comprehensive treatment benefit–harm profiles.

In addition, the framework explicitly balances two central aspects of decision-making across outcomes: the overall and the worst-case treatment performance. The VIKOR algorithm achieves this compromise through the $Q$ metric, which ranks treatments according to their proximity to an ideal solution. This compromise between overall and worst-case treatment performance renders VIKOR more robust than other methods, such as spie charts, in avoiding the obscuration of differences between efficacy and safety outcomes when these are in conflict. This was illustrated in the reanalysis of the psoriasis data[17,37] presented in the Supplementary Material. By imposing explicit mathematical decision rules, VIKOR identifies treatments that achieve an optimal balance between overall and worst-case performance, thereby supporting HTA and decision-making. Although specifying outcome-specific weights may increase practical complexity, this feature allows the framework to incorporate patient preferences and priorities regarding outcomes, enabling more personalized decision-making. This aligns with recommendations from the Institute of Medicine[46] and the Agency for Healthcare Research and Quality[47] emphasizing the importance of integrating patient preferences into benefit–harm assessments.

Despite these advantages, the proposed framework has limitations. mvNMA introduces additional modelling complexity, particularly as the number of outcomes increases, and gains in precision may be modest when evidence is complete across outcomes[5]. Moreover, the improved precision and full estimation of treatment effects under the mvNMA(DM) approach rely on the assumption of constant relative potency across outcomes. Assessing the plausibility of this assumption can be challenging, as it requires interpretation of a variance component, which may be difficult for non-statistically trained audiences. In addition, when treatments are missing not at random across outcomes, borrowing information may introduce bias rather than strength[48], in such cases fitting separate outcome-specific NMA models may yield similar results with substantially lower complexity. To enable the use of VIKOR in such cases, we further implemented the algorithm within the R package **netmeta**[38].

Limitations also arise conceptually as the construction of treatment hierarchies across multiple outcomes inevitably carries the risk of obscuring harms in favour of benefits, or vice versa. Although the VIKOR algorithm was shown to be robust and to avoid such obscurations, results should be interpreted alongside the underlying treatment effect estimates and outcome-specific hierarchies rather than in isolation. Furthermore, the set of compromise treatments identified by VIKOR should not be interpreted as prescriptive recommendations. In practice, HTA bodies and guideline developers consider benefit–harm profiles in conjunction with additional factors, including costs and broader social, ethical, and environmental considerations. Consequently, the proposed framework should be viewed as a transparent decision-support tool that supports, rather than replaces, the wider evaluative process.

Overall, the proposed framework offers an alternative strategy for assessing treatment benefit–harm profiles across multiple outcomes. It combines MCDA algorithms with the modelling capacities of mvNMA to support the construction of across-outcomes treatment hierarchies. The accompanying R package, **mvnma**, implements mvNMA model fitting, computes outcome-specific treatment hierarchies, and enables application of the VIKOR and fuzzy-VIKOR algorithms in a user-friendly manner. We expect this package to encourage wider adoption of multivariate approaches in applied clinical and policy settings. To our knowledge, this is the first R package specifically developed for mvNMA. Future work could extend the framework by incorporating clinically meaningful effect thresholds[49] into outcome-specific ranking metrics and by evaluating the extrapolation properties of the mvNMA(DM) model under different patterns of missing outcome data. In addition, extending mvNMA to the meta-regression[50,51] would allow the model to account for treatment-by-covariate interactions and to estimate adjusted treatment effects, thereby refining treatment hierarchies both within and across outcomes. Investigators can use the proposed framework as a primary analytical approach or as a complementary sensitivity analysis alongside established methods, such as outcome-specific NMAs or alternative ranking approaches including mP-scores.

**Acknowledgments:** We thank Prof. Christopher Schmid for his valuable comments and feedback on the first version of this manuscript. Theodoros Evrenoglou was supported by the Deutsche Forschungsgemeinschaft (DFG, German Research Foundation) under the Project ID-554095932.

**Conflicts of interest statement**: None

**Data availability statement**: All codes and data used to produce the results of this manuscript are open and freely available in the following GitHub repository: https://github.com/TEvrenoglou/mvnma_article.

# Supplementary Material: Identifying compromise solutions across multiple treatment benefit–harm profiles in multivariate network meta-analysis

## The fully hierarchical mvNMA model and its connection to the hybrid mvNMA model

For simplicity, we present the standard fully-hierarchical mvNMA model and explain its connection to the hybrid mvNMA model described in the main manuscript, for the case of two outcomes and two-arm studies. Specifically, suppose that $J = 2$ outcomes are available from each of $N$ studies comparing a set of $K$ treatments. Following the notation of the main manuscript, a fully hierarchical mvNMA model can be written as,

$$y_{ij,k} \sim \mathrm{N}\left(\delta_{ij,k}, \mathbf{S_i}\right)$$

$$\delta_{ij,k} \sim \mathrm{N}(\theta_{jb,k}, \mathbf{\Sigma_i})$$

where $\mathbf{S_i}$ denotes the within-study variance-covariance matrix, which has the following form,

$$\mathbf{S}_i = \begin{pmatrix} s_{i1,k}^2 & \rho_{i,12}^W s_{i1,k} s_{i2,k} \\ \rho_{i,12}^W s_{i1,k} s_{i2,k} & s_{i2,k}^2 \end{pmatrix}$$

with $\rho_{i,12}^W$ denoting the within-study outcome correlation and indicating the association between point estimates $y_{i1,k}$ and $y_{i2,k}$ in study $i$. Note that in this model the quantity $\rho_{i,12}^W$ is assumed to be known and extracted from each study $i = 1,2, \dots N$. Matrix $\mathbf{\Sigma_i}$ denotes the across-study variance-covariance matrix, which has the following form,

$$\mathbf{\Sigma}_i = \begin{pmatrix} \tau_1^2 & \rho_{12}^B \tau_1 \tau_2 \\ \rho_{12}^B \tau_1 \tau_2 & \tau_2^2 \end{pmatrix}$$

with $\tau_j^2$ indicating between study variation in terms of outcome $j = 1,2$ and $\rho_{12}^B$ denoting the between-study correlation, which indicates how the underlying true treatment effects $\theta_{jb,k}$ are related across the $N$ studies. For the case of the fully-hierarchical model, inference typically relies on the marginal model, which can be written as,

$$y_{ij,k} \sim \mathrm{N}\left(\theta_{jb,k}, \mathbf{V_i}\right)$$

where $\mathbf{V_i}$ denotes the overall variance-covariance matrix, which has the following form,

$$\mathbf{V}_i = \begin{pmatrix} s_{i1,k}^2 + \tau_1^2 & \rho_{i,12}^W s_{i1,k} s_{i2,k} + \rho_{12}^B \tau_1 \tau_2 \\ \rho_{i,12}^W s_{i1,k} s_{i2,k} + \rho_{12}^B \tau_1 \tau_2 & s_{i2,k}^2 + \tau_2^2 \end{pmatrix}$$

The previous mvNMA model partitions both overall variation and overall correlation into within- and across-study components using a fully hierarchical structure. However, study-specific correlation terms $\rho_{i,12}^W$ are seldom reported at the study level, complicating the standard mvNMA model fit. Alternatively, following previous work[1,2], we consider a hybrid mvNMA model that is written as

$$y_{ij,k} \sim \mathrm{N}\left(\theta_{jb,k}, \mathbf{\Sigma_i}\right)$$

where now $\mathbf{\Sigma_i}$ has the form

$$\mathbf{\Sigma}_i = \begin{pmatrix} \zeta_{i1,k} & \rho_{12}\sqrt{\zeta_{i1,k}\zeta_{i2,k}} \\ \rho_{12}\sqrt{\zeta_{i1,k}\zeta_{i2,k}} & \zeta_{i2,k} \end{pmatrix}$$

with $\zeta_{i1,k} = s_{i1,k}^2 + \psi_1^2$ and $\zeta_{i2,k} = s_{i2,k}^2 + \psi_2^2$. The core difference between the fully hierarchical and the hybrid mvNMA model is that the latter partitions only the overall variation into within- and across-study variance and models the overall correlation directly, without any partition[2]. Consequently, the hybrid mvNMA model does not rely on the extraction of the within-study correlation terms and estimates the outcome correlation $\rho_{12}$ as an amalgam of within and between study correlations $\rho_{i,12}^W$ and $\rho_{12}^B$. Note that the between study variance terms $\tau_j^2$ and $\psi_j^2$ are denoted here with different symbols, as these are not always equivalent. This is because, by not partitioning the overall correlation, the hybrid model does not have a fully hierarchical structure[2].

Overall, the standard fully-hierarchical structure provides a more standard modelling framework but relies on the availability of within-study correlation terms, which are seldom reported, complicating the practical use of this model. In contrast, the hybrid mvNMA model simplifies this by modelling the overall between-outcome correlation as an amalgam of within- and between-study correlation terms. This comes at the cost of losing the fully-hierarchical model structure, which is the standard modelling framework in the meta-analytic field. Previous simulation studies compared these two approaches and concluded that they yield very similar treatment effect estimates, with differences existing only in the limiting cases of $\hat{\rho}_{12} = 1$ or $\hat{\rho}_{12} = -1$, which favored the fully hierarchical approach[2].

### Deriving the variance-covariance matrix $\mathbf{\Sigma_i}$ of the hybrid mvNMA model

Following the considerations outlined in the main manuscript, we provide expressions for the overall variance–covariance matrix $\mathbf{\Sigma_i}$ under the hybrid mvNMA model. This formulation assumes that all pairwise comparisons share (i) a common outcome-specific heterogeneity parameter and (ii) common between-outcome correlations. Using the same notation of the main manuscript, the variance-covariance matrix $\mathbf{\Sigma_i}$ for two outcomes and a three-arm study takes the following general form:

$$\mathbf{\Sigma}_i = \begin{pmatrix} \zeta_{i1,k} & \cdot & \cdot & \cdot \\ \rho_{12}\sqrt{\zeta_{i1,k}\zeta_{i2,k}} & \zeta_{i2,k} & \cdot & \cdot \\ x_1\sqrt{\zeta_{i1,k}\zeta_{i1,l}} & x_2\sqrt{\zeta_{i2,k}\zeta_{i1,l}} & \zeta_{i1,l} & \cdot \\ x_2\sqrt{\zeta_{i1,k}\zeta_{i2,l}} & x_1\sqrt{\zeta_{i2,k}\zeta_{i2,l}} & \rho_{12}\sqrt{\zeta_{i1,l}\zeta_{i2,l}} & \zeta_{i2,l} \end{pmatrix}$$

where the diagonal elements $\zeta_{ij,k}$ and $\zeta_{ij,l}$ denote the overall variation of the contrast $kb$ and $lb$, respectively, for outcome $j = 1,2$. The off-diagonal elements comprise three types of covariance. The first arises when the same contrast (e.g. $kb$) is evaluated across different outcomes. Letting $\rho_{12}$ denote the correlation between outcomes 1 and 2, this covariance is obtained simply as: $\text{Cov}(y_{i1,k}, y_{i2,k}) = \rho_{12}\sqrt{\zeta_{i1,k}\zeta_{i2,k}}$.

The second type of covariance reflects the dependence induced by the multi-arm designs, whereby two distinct contrasts (e.g. $kb$ and $lb$) are evaluated for the same outcome $j$. This quantity is equal to $\text{Cov}(y_{i1,k}, y_{i1,l}) = x_1\sqrt{\zeta_{i1,k}\zeta_{i1,l}}$. To derive $x_1$ we note that

$$\text{Var}(y_{ij,lb}) = \text{Var}(y_{ij,lk} + y_{ij,kb}) = \text{Var}(y_{ij,lk}) + \text{Var}(y_{ij,kb}) - 2\text{Cov}(y_{ij,lk}, y_{ij,kb}).$$

Assuming that $\sigma^2_{ij,kb} = \sigma^2_{ij,lb} = \sigma^2_{ij,kl} = \sigma^2_j \Rightarrow \zeta_{ij,kb} = \zeta_{ij,kl} = \zeta_{ij,lb} = \zeta_j$ the above Equation becomes

$$\zeta_j = 2\zeta_j - 2x_1\zeta_j,$$

which yields $x_1 = \frac{1}{2}$.

The third type of covariance arises when different contrasts are evaluated across different outcomes. This quantity is given by $\text{Cov}(y_{i1,k}, y_{i2,l}) = x_2\sqrt{\zeta_{i1,k}\zeta_{i2,l}}$. To derive $x_2$ we note that

$$\mathrm{Var}(y_{i2,lb} - y_{i1,kb}) = \mathrm{Var}(y_{i2,lb}) + \mathrm{Var}(y_{i1,kb}) - 2\mathrm{Cov}(y_{i2,lb}, y_{i1,kb})$$
$$= \zeta_{i1,k} + \zeta_{i2,l} - 2x_2\sqrt{\zeta_{i1,k}\zeta_{i2,l}}$$

Again, assuming common variances within each outcome, that is $\zeta_{ij,kb} = \zeta_{ij,kl} = \zeta_{ij,lb} = \zeta_j$ this relation becomes

$$\mathrm{Var}(y_{i2,lb} - y_{i1,kb}) = \zeta_1 + \zeta_2 - 2x_2\sqrt{\zeta_1\zeta_2}$$

Alternatively, consistency implies $y_{i2,lb} = \ y_{i2,lk} + y_{i2,kb}$, and therefore

$$\mathrm{Var}(y_{i2,lb} - y_{i1,kb}) = \mathrm{Var}(y_{i2,lk} + y_{i2,kb} - y_{i1,kb}) =$$

$$\mathrm{Var}(y_{i2,lk}) + \mathrm{Var}(y_{i2,kb}) + \mathrm{Var}(y_{i1,kb}) - 2\mathrm{Cov}(y_{i2,lk}, y_{i1,kb}) - 2\mathrm{Cov}(y_{i2,kb}, y_{i1,kb})$$
$$+ 2\mathrm{Cov}(y_{i2,lk}, y_{i2,kb})$$
$$= \mathrm{Var}(y_{i2,lk}) + \mathrm{Var}(y_{i2,kb}) + \mathrm{Var}(y_{i1,kb}) + 2\mathrm{Cov}(y_{i2,kl}, y_{i1,kb})$$
$$- 2\mathrm{Cov}(y_{i2,kb}, y_{i1,kb}) - 2\mathrm{Cov}(y_{i2,kl}, y_{i2,kb})$$
$$= \zeta_2 + \zeta_2 + \zeta_1 + 2x_2\sqrt{\zeta_1\zeta_2} - 2\rho_{12}\sqrt{\zeta_1\zeta_2} - 2\frac{1}{2}\zeta_2$$
$$= \zeta_1 + \zeta_2 - 2\rho_{12}\sqrt{\zeta_1\zeta_2} + 2x_2\sqrt{\zeta_1\zeta_2}$$

Equating the two expressions of $\mathrm{Var}(\mathrm{y}_{\mathrm{i2,lb}} - y_{i1,kb})$ gives

$$\zeta_1 + \zeta_2 - 2x_2\sqrt{\zeta_1\zeta_2} = \zeta_1 + \zeta_2 - 2\rho_{12}\sqrt{\zeta_1\zeta_2} + 2x_2\sqrt{\zeta_1\zeta_2} \Rightarrow x_2 = \frac{\rho_{12}}{2}.$$

Substituting $x_1$ and $x_2$ into the generic form of $\mathbf{\Sigma_i}$ yields the following matrix for a three-arm study with two outcomes,

$$\mathbf{\Sigma}_i = \begin{pmatrix} \zeta_{i1,k} & \cdot & \cdot & \cdot \\ \rho_{12}\sqrt{\zeta_{i1,k}\zeta_{i2,k}} & \zeta_{i2,k} & \cdot & \cdot \\ \frac{1}{2}\sqrt{\zeta_{i1,k}\zeta_{i1,l}} & \frac{\rho_{12}}{2}\sqrt{\zeta_{i2,k}\zeta_{i1,l}} & \zeta_{i1,l} & \cdot \\ \frac{\rho_{12}}{2}\sqrt{\zeta_{i1,k}\zeta_{i2,l}} & \frac{1}{2}\sqrt{\zeta_{i2,k}\zeta_{i2,l}} & \rho_{12}\sqrt{\zeta_{i1,l}\zeta_{i2,l}} & \zeta_{i2,l} \end{pmatrix}$$

Overall, in this matrix, terms of the form $\rho_{12}\sqrt{\zeta_{i1,k}\zeta_{i2,k}}$ represent the covariance between treatment effects for the same comparison across two different outcomes. As in NMA, terms of the form $\frac{1}{2}\sqrt{\zeta_{i1,k}\zeta_{i1,l}}$ capture outcome-specific covariance induced by treatment effects sharing a common study reference in multi-arm studies. Finally, terms of the form $\frac{\rho_{12}}{2}\sqrt{\zeta_{i1,k}\zeta_{i2,l}}$ represent the covariance between treatment effects for treatments $k$ and $l$, compared against a common reference treatment group $b$ across two different outcomes. Following the same rationale, $\mathbf{\Sigma_i}$ can be constructed for any $K_i$-arm study reporting $J$ outcomes. As an illustration, a more complex form of $\mathbf{\Sigma_i}$, for a three-arm study with five outcomes, is given below.

$$
\mathbf{\Sigma}_i = \begin{pmatrix}
\zeta_{i1,k} & \cdot & \cdot & \cdot & \cdot & \cdot & \cdot & \cdot & \cdot & \cdot \\
\rho_{12}\sqrt{\zeta_{i1,k}\zeta_{i2,k}} & \zeta_{i2,k} & \cdot & \cdot & \cdot & \cdot & \cdot & \cdot & \cdot & \cdot \\
\rho_{13}\sqrt{\zeta_{i1,k}\zeta_{i3,k}} & \rho_{23}\sqrt{\zeta_{i2,k}\zeta_{i3,k}} & \zeta_{i3,\kappa} & \cdot & \cdot & \cdot & \cdot & \cdot & \cdot & \cdot \\
\rho_{14}\sqrt{\zeta_{i1,k}\zeta_{i4,k}} & \rho_{24}\sqrt{\zeta_{i2,k}\zeta_{i4,k}} & \rho_{34}\sqrt{\zeta_{i3,k}\zeta_{i4,k}} & \zeta_{i4,\kappa} & \cdot & \cdot & \cdot & \cdot & \cdot & \cdot \\
\rho_{15}\sqrt{\zeta_{i1,k}\zeta_{i5,k}} & \rho_{25}\sqrt{\zeta_{i2,k}\zeta_{i5,k}} & \rho_{35}\sqrt{\zeta_{i3,k}\zeta_{i5,k}} & \rho_{45}\sqrt{\zeta_{i4,k}\zeta_{i5,k}} & \zeta_{i5,\kappa} & \cdot & \cdot & \cdot & \cdot & \cdot \\
\frac{1}{2}\sqrt{\zeta_{i1,k}\zeta_{i1,l}} & \frac{\rho_{12}}{2}\sqrt{\zeta_{i2,k}\zeta_{i1,l}} & \frac{\rho_{13}}{2}\sqrt{\zeta_{i3,k}\zeta_{i1,l}} & \frac{\rho_{14}}{2}\sqrt{\zeta_{i4,k}\zeta_{i1,l}} & \frac{\rho_{15}}{2}\sqrt{\zeta_{i5,k}\zeta_{i1,l}} & \zeta_{i1,l} & \cdot & \cdot & \cdot & \cdot \\
\frac{\rho_{12}}{2}\sqrt{\zeta_{i1,k}\zeta_{i2,l}} & \frac{1}{2}\sqrt{\zeta_{i2,k}\zeta_{i2,l}} & \frac{\rho_{23}}{2}\sqrt{\zeta_{i3,k}\zeta_{i2,l}} & \frac{\rho_{24}}{2}\sqrt{\zeta_{i4,k}\zeta_{i2,l}} & \frac{\rho_{25}}{2}\sqrt{\zeta_{i5,k}\zeta_{i2,l}} & \rho_{12}\sqrt{\zeta_{i1,l}\zeta_{i2,l}} & \zeta_{i2,l} & \cdot & \cdot & \cdot \\
\frac{\rho_{13}}{2}\sqrt{\zeta_{i1,k}\zeta_{i3,l}} & \frac{\rho_{23}}{2}\sqrt{\zeta_{i2,k}\zeta_{i3,l}} & \frac{1}{2}\sqrt{\zeta_{i3,k}\zeta_{i3,l}} & \frac{\rho_{34}}{2}\sqrt{\zeta_{i4,k}\zeta_{i3,l}} & \frac{\rho_{35}}{2}\sqrt{\zeta_{i5,k}\zeta_{i3,l}} & \rho_{13}\sqrt{\zeta_{i1,l}\zeta_{i3,l}} & \rho_{23}\sqrt{\zeta_{i2,l}\zeta_{i3,l}} & \zeta_{i3,bl} & \cdot & \cdot \\
\frac{\rho_{14}}{2}\sqrt{\zeta_{i1,k}\zeta_{i4,l}} & \frac{\rho_{24}}{2}\sqrt{\zeta_{i2,k}\zeta_{i4,l}} & \frac{\rho_{34}}{2}\sqrt{\zeta_{i3,k}\zeta_{i4,l}} & \frac{1}{2}\sqrt{\zeta_{i4,k}\zeta_{i4,l}} & \frac{\rho_{45}}{2}\sqrt{\zeta_{i5,k}\zeta_{i4,l}} & \rho_{14}\sqrt{\zeta_{i1,l}\zeta_{i4,l}} & \rho_{24}\sqrt{\zeta_{i2,l}\zeta_{i4,l}} & \rho_{34}\sqrt{\zeta_{i3,l}\zeta_{i4,l}} & \zeta_{i4,l} & \cdot \\
\frac{\rho_{15}}{2}\sqrt{\zeta_{i1,k}\zeta_{i5,l}} & \frac{\rho_{25}}{2}\sqrt{\zeta_{i2,k}\zeta_{i5,l}} & \frac{\rho_{35}}{2}\sqrt{\zeta_{i3,k}\zeta_{i5,l}} & \frac{\rho_{45}}{2}\sqrt{\zeta_{i4,k}\zeta_{i5,l}} & \frac{1}{2}\sqrt{\zeta_{i5,k}\zeta_{i5,l}} & \rho_{15}\sqrt{\zeta_{i1,l}\zeta_{i5,l}} & \rho_{25}\sqrt{\zeta_{i2,l}\zeta_{i5,l}} & \rho_{35}\sqrt{\zeta_{i3,l}\zeta_{i5,l}} & \rho_{45}\sqrt{\zeta_{i4,l}\zeta_{i5,l}} & \zeta_{i5,l}
\end{pmatrix}
$$

### Fuzzy VIKOR

The fuzzy VIKOR algorithm[3] extends the standard VIKOR approach[4,5] presented in the main manuscript to accommodate outcome-specific ranking metrics that are reported with uncertainty. Let $m_{jk}$ denote the median rank of treatment $k = 1,2,\dots,K$ for outcome $j = 1,2,\dots,K$ and let $l_{jk}$ and $u_{jk}$ denote the corresponding lower and upper bounds of the 95% credible interval (CrI) for $m_{jk}$. Following the main manuscript, and for mathematical convenience, we assume complete evidence for all treatments across all outcomes of interest.

As in standard VIKOR, the first step of the fuzzy VIKOR algorithm involves defining the positive ideal solution. For median ranks, the positive ideal solution, is set equal to 1, corresponding to the performance of a hypothetical treatment achieving the first position in the treatment hierarchy across all outcomes. A key distinction relative to probabilistic ranking metrics (e.g., SUCRA) is that, for median ranks, the positive ideal solution corresponds to the smallest attainable value of the metric, whereas for probabilistic metrics it corresponds to the largest. Consequently, overall and worst-case treatment performance can be defined for each of the elements $l_{jk}, m_{jk}$ and $u_{jk}$ by reversing the signs in formulas (8) and (9) of the main manuscript. Specifically, let

$$\tilde{d}_{jk} = \left(\frac{l_{jk} - 1}{K - 1}, \frac{m_{jk} - 1}{K - 1}, \frac{u_{jk} - 1}{K - 1}\right) \tag{11}$$

denote the normalized differences of each median-rank element from the positive ideal solution of 1. In Equation (11) the denominator $K - 1$ is used as a normalizing constant, ensuring that all elements of $\tilde{d}_{jk}$ lie between 0 and 1. This step is redundant for probabilistic ranking metrics in the standard VIKOR framework, where the negative ideal solution equals 0 and the denominator cancels out.

The elements of the fuzzy vector summarizing overall treatment performance across all outcomes for treatment, denoted by $\tilde{S}_k = \left(S_k^l, S_k^m, S_k^u\right)$, are then obtained as

$$\tilde{S}_k = \left(\sum_j w_j \tilde{d}_{jk}^{(l)}, \sum_j w_j \tilde{d}_{jk}^{(m)}, \sum_j w_j \tilde{d}_{jk}^{(u)}\right) \tag{12}$$

where $w_j, j = 1,2,\ldots,J$ denotes the weight assigned to outcome $j$, and $\tilde{d}_{jk}^{(q)}$ denotes the $q^{th}$ element of vector $\tilde{d}_{jk}$ with $q = 1,2,3$. Similarly, the fuzzy worst-case performance of treatment $k$ across all outcomes, denoted by $\tilde{R}_k = \left(R_k^l, R_k^m, R_k^u\right)$, is given as

$$\tilde{R}_k = \left(\max_j\left[w_j\tilde{d}_{jk}^{(l)}\right], \max_j\left[w_j\tilde{d}_{jk}^{(m)}\right], \max_j\left[w_j\tilde{d}_{jk}^{(u)}\right]\right) \tag{13}$$

Given the fuzzy overall and worst-case performance for each treatment $k$, we define the $q^{th}$ element of the fuzzy compromise metric $\tilde{Q}_k = \left(Q_k^l, Q_k^m, Q_k^u\right)$ as

$$\tilde{Q}_k^{(q)} = v\,\frac{\tilde{S}_k^{(q)} - \min\left[\tilde{S}_k\right]}{\max\left[\tilde{S}_k^{(u)}\right] - \min\left[\tilde{S}_k^{(l)}\right]} + (1-v)\,\frac{\tilde{R}_k^{(q)} - \tilde{R}^*}{\max\left[\tilde{R}_k^{(u)}\right] - \min\left[\tilde{R}_k^{(l)}\right]} \tag{14}$$

Following Opricovic et al.[3], the defuzzification of fuzzy triplets $\tilde{S}_k$, $\tilde{R}_k$ and $\tilde{Q}_k$ is performed using the “second weighted mean” method, defined through the mapping

$$f: R^{3+} \to R, \qquad f(x,y,z) = \frac{2y+x+z}{4} \tag{15}$$

Applying Equation (5) to $\tilde{S}_k$, $\tilde{R}_k$ and $\tilde{Q}_k$ yields the scalar quantities $S_k$, $R_k$ and $Q_k$, respectively. These correspond to the standard VIKOR ranking metrics $S$, $R$ and $Q$. The remaining steps of the fuzzy VIKOR algorithm then coincide with those of the standard VIKOR procedure described in the main manuscript.

## Additional results for the physical therapy treatments network

### Convergence diagnostics

**S-Table 1:** Convergence diagnostics in terms of the R-hat statistic for all basic estimates in the physical therapy treatments network. For each parameter estimate, the effective sample size is shown in square brackets.

| **Comparison** *Treatment vs No Treatment* | **Pain** | | **Disability** | |
|---|---|---|---|---|
| | R-hat [n.eff] mvNMA(standard) | R-hat [n.eff] mvNMA(DM) | R-hat [n.eff] mvNMA(standard) | R-hat [n.eff] mvNMA(DM) |
| Aerobic exercise | 1.00 [2600] | 1.00 [1300] | 1.00 [2900] | 1.00 [1200] |
| Aquatic exercise | 1.00 [3100] | 1.00 [1000] | 1.00 [4000] | 1.00 [1500] |
| Diathermy (high) | 1.00 [4000] | 1.00 [4000] | 1.00 [2100] | 1.00 [2600] |
| Diathermy (low) | 1.00 [4000] | 1.00 [4000] | 1.00 [3000] | 1.00 [4000] |
| Education | 1.00 [2200] | 1.00 [680] | 1.00 [2300] | 1.00 [1000] |
| Electrical stimulation | 1.00 [1400] | 1.00 [1000] | 1.00 [4000] | 1.00 [1500] |
| Placebo | NA | NA | NA | NA |
| Proprioception exercise | 1.00 [4000] | 1.00 [2300] | 1.00 [2700] | 1.00 [1000] |
| Strength exercise | 1.00 [4000] | 1.00 [2500] | 1.00 [2600] | 1.00 [1700] |
| Ultrasound | 1.00 [4000] | 1.00 [4000] | 1.00 [4000] | 1.00 [2600] |

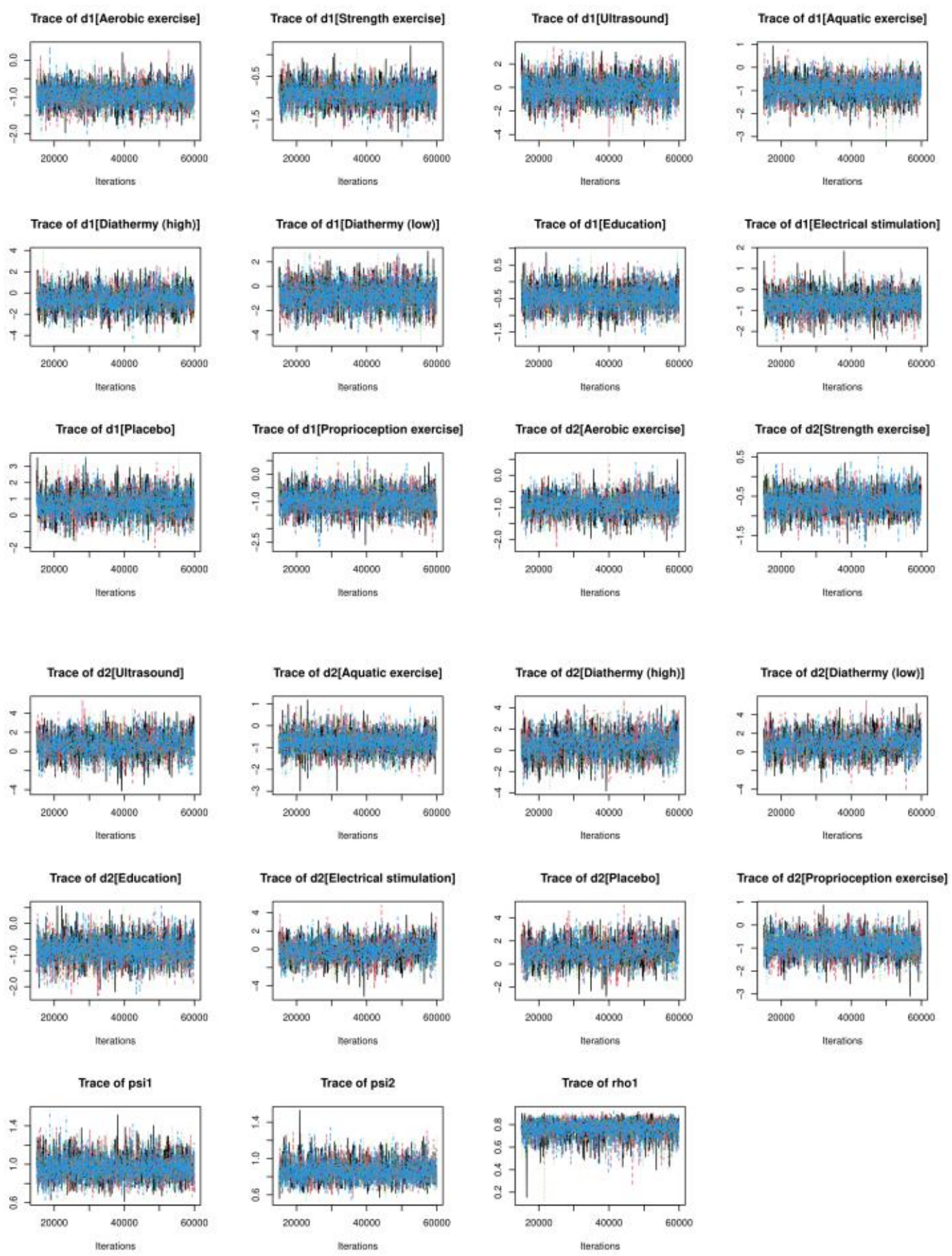

**S-Figure 1:** Traceplots for parameters of interest in the physical therapy treatments network for the mvNMA(standard) model. Parameters $d$, $\psi$ (psi) and $\rho$ (rho) correspond to the MCMC chains for basic treatment effects, outcome-specific heterogeneity, and outcome correlation, respectively. The numbers in each trace plot title denote the outcomes, with 1 representing pain and 2 representing disability.

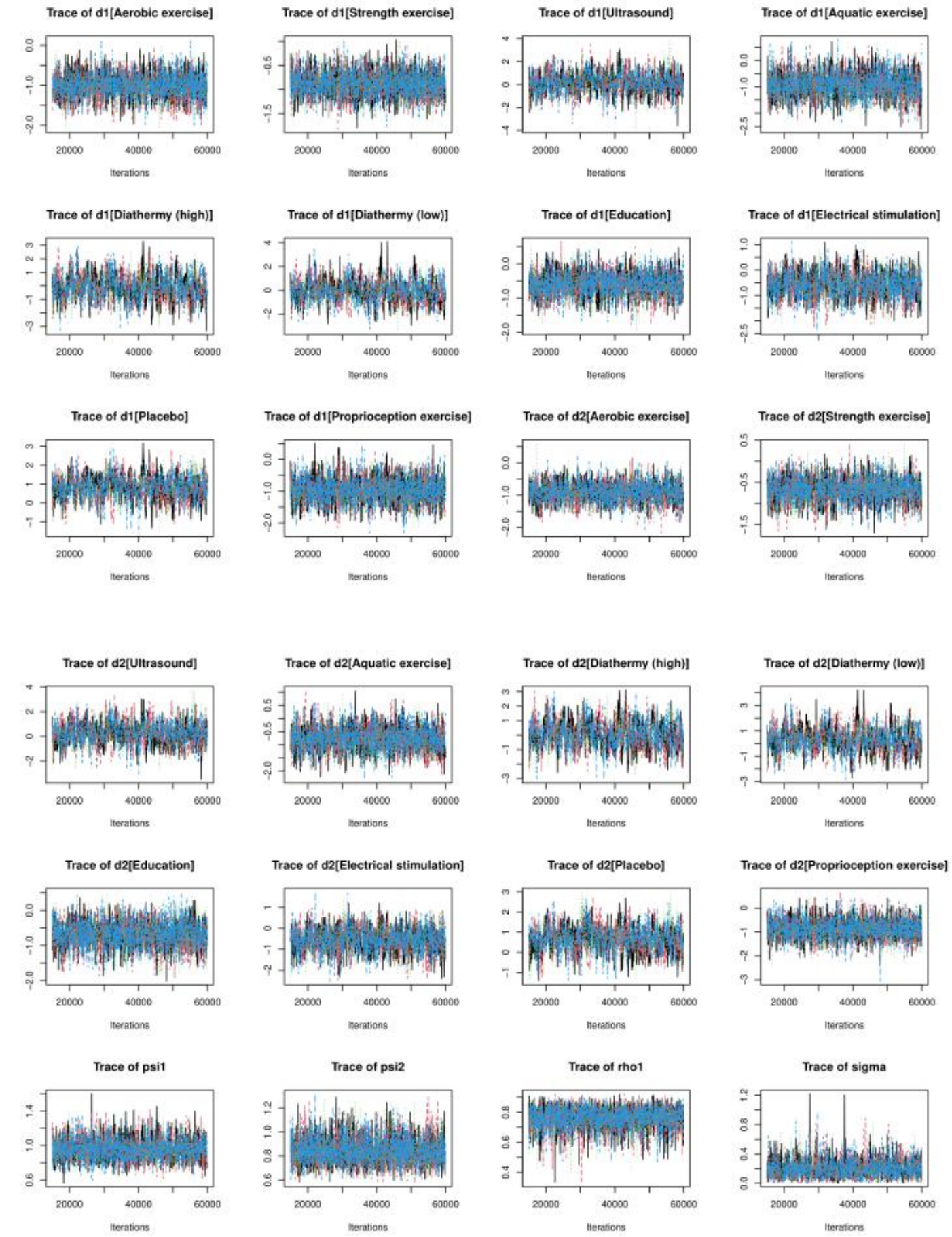

**S-Figure 2:** Traceplots for parameters of interest in the physical therapy treatments network for the mvNMA(DM) model. Parameters $d$, $\psi$ (psi) and $\rho$ (rho) correspond to the MCMC chains for basic treatment effects, outcome-specific heterogeneity, and outcome correlation, respectively. The numbers in each trace plot title denote the outcomes, with 1 representing pain and 2 representing disability.

### Outcome-specific treatment hierarchies using median ranks

**S-Table 2:** Outcome-specific median ranks and 95% CrIs for both the Pain and Disability outcomes, from the mvNMA(standard) and mvNMA(DM) models.

| Treatment | Pain | | Disability | |
|---|---|---|---|---|
| | Median rank [95% CrI] mvNMA(standard) | Median rank [95% CrI] mvNMA(DM) | Median rank [95% CrI] mvNMA(standard) | Median rank [95% CrI] mvNMA(DM) |
| Aerobic exercise | **4 [1, 8]** | **3 [1, 7]** | **3 [1, 7]** | **3 [1, 7]** |
| Proprioception exercise | **4 [1, 9]** | **3 [1, 8]** | **3 [1, 9]** | **3 [1, 9]** |
| Strength exercise | **4 [1, 8]** | **4 [1, 8]** | 4 [1, 9] | **4 [1, 8]** |
| Aquatic exercise | 4 [1, 10] | 3 [1, 9] | 4 [1, 10] | 4 [1, 9] |
| Diathermy (low) | 5 [1, 10] | 8 [1, 11] | 9 [2, 11] | 9 [1, 11] |
| Electrical stimulation | 6 [2, 9] | 6 [1, 9] | 7 [1, 10] | 5 [1, 9] |
| Diathermy (high) | 6 [1, 11] | 8 [1, 11] | 8 [1, 11] | 8 [1, 11] |
| Education | 7 [3, 10] | 5 [2, 9] | **3 [1, 8]** | 4 [1, 9] |
| Ultrasound | 8 [1, 11] | 9 [1, 11] | 9 [2, 11] | 9 [1, 11] |
| No treatment | 9 [6, 11] | 8 [6, 11] | 7 [5, 11] | 8 [5, 11] |
| Placebo | 11 [7, 11] | 11 [9, 11] | 10 [5, 11] | 10 [7, 11] |

### Results of fuzzy VIKOR

**S-Table 3:** Fuzzy VIKOR results for median ranks obtained from the mvNMA(standard) model.

| Treatment | Q | S | R |
|---|---|---|---|
| Aerobic exercise | **0.00** | 0.24 | 0.14 |
| Proprioception exercise | **0.02** | 0.27 | 0.14 |
| Aquatic exercise | **0.08** | 0.31 | 0.16 |
| Strength exercise | 0.12 | 0.34 | 0.18 |
| Electrical stimulation | 0.31 | 0.46 | 0.23 |
| Education | 0.35 | 0.41 | 0.29 |
| Diathermy (high) | 0.53 | 0.57 | 0.33 |
| Diathermy (low) | 0.64 | 0.59 | 0.39 |
| Ultrasound | 0.67 | 0.68 | 0.36 |
| No treatment | 0.75 | 0.72 | 0.40 |
| Placebo | 1.00 | 0.91 | 0.47 |

**S-Table 4:** Fuzzy VIKOR results for median ranks obtained from mvNMA(DM) model.

| Treatment | Q | S | R |
|---|---|---|---|
| Aerobic exercise | **0.00** | 0.08 | 0.04 |
| Proprioception exercise | **0.04** | 0.10 | 0.05 |
| Strength exercise | **0.07** | 0.11 | 0.05 |
| Aquatic exercise | 0.09 | 0.11 | 0.06 |
| Education | 0.16 | 0.13 | 0.07 |
| Electrical stimulation | 0.19 | 0.14 | 0.07 |
| Diathermy (high) | 0.35 | 0.20 | 0.10 |
| Diathermy (low) | 0.39 | 0.21 | 0.11 |
| Ultrasound | 0.40 | 0.22 | 0.11 |
| No treatment | 0.47 | 0.24 | 0.12 |
| Placebo | 0.68 | 0.30 | 0.16 |

## Additional results for the antidepressants dataset

**S-Table 5:** Outcome-specific heterogeneity estimates from both mvNMA and independent NMA models, across all outcomes in the antidepressants dataset.

| Outcome | mvNMA(standard) ($\psi$) | mvNMA(DM) ($\psi$) | NMA ($\tau$) |
|---|---|---|---|
| Early Response | 0.20 | 0.18 | 0.19 |
| Early Remission | 0.09 | 0.09 | 0.15 |
| Adverse Events | 0.30 | 0.29 | 0.27 |
| Loss to follow up | 0.22 | 0.21 | 0.26 |
| Loss to follow up due to adverse events | 0.20 | 0.17 | 0.20 |

## Convergence diagnostics

**S-Table 6:** Convergence diagnostics in terms of the R-hat statistic for all basic estimates obtained from the mvNMA(standard) model in the antidepressants dataset. For each parameter, the effective sample size appears in square brackets.

| **Comparison** *Treatment vs Placebo* | **Early Response** | **Early Remission** | **Adverse Events** | **Loss to follow up** | **Loss to follow up due to AE** |
|---|---|---|---|---|---|
| Hypericum | 1.00 [2200] | 1.00 [4000] | 1.00 [4000] | 1.00 [1600] | 1.00 [3900] |
| Low-dose SARI | 1.00 [3300] | 1.00 [2000] | 1.00 [4000] | 1.00 [3800] | 1.00 [4000] |
| NaSSa | 1.00 [4000] | 1.00 [4000] | 1.00 [4000] | 1.00 [4000] | 1.01 [4000] |
| NRI | 1.00 [4000] | 1.00 [4000] | NA | 1.00 [4000] | 1.00 [4000] |
| Placebo | NA | NA | NA | NA | NA |
| rMAO-A | 1.00 [1900] | 1.00 [3100] | 1.00 [4000] | 1.00 [4000] | 1.00 [4000] |
| SNRI | 1.00 [2400] | 1.00 [4000] | 1.00 [3700] | 1.00 [880] | 1.00 [3900] |
| SSRI | 1.00 [2800] | 1.00 [4000] | 1.00 [4000] | 1.00 [2900] | 1.00 [4000] |
| TCA | 1.00 [4000] | 1.00 [4000] | 1.00 [2700] | 1.00 [4000] | 1.00 [4000] |

**S-Table 7:** Convergence diagnostics in terms of the R-hat statistic for all basic estimates obtained from the mvNMA(DM) model in the antidepressants dataset. For each parameter, the effective sample size appears in square brackets.

| **Comparison** *Treatment vs Placebo* | **Early Response** | **Early Remission** | **Adverse Events** | **Loss to follow up** | **Loss to follow up due to AE** |
|---|---|---|---|---|---|
| Hypericum | 1.01 [300] | 1.01 [470] | 1.00 [4000] | 1.00 [1500] | 1.00 [4000] |
| Low-dose SARI | 1.00 [1500] | 1.00 [2700] | 1.00 [3000] | 1.00 [2600] | 1.00 [4000] |
| NaSSa | 1.00 [1000] | 1.00 [1800] | 1.00 [2200] | 1.00 [4000] | 1.00 [4000] |
| NRI | 1.00 [1500] | 1.00 [4000] | 1.00 [4000] | 1.00 [4000] | 1.00 [4000] |
| Placebo | NA | NA | NA | NA | NA |
| rMAO-A | 1.00 [710] | 1.00 [1600] | 1.00 [4000] | 1.00 [1800] | 1.00 [3300] |
| SNRI | 1.00 [1900] | 1.00 [4000] | 1.00 [2400] | 1.00 [2800] | 1.00 [4000] |
| SSRI | 1.00 [840] | 1.00 [1800] | 1.00 [4000] | 1.00 [1000] | 1.00 [3900] |
| TCA | 1.01 [460] | 1.00 [1200] | 1.00 [2000] | 1.00 [1900] | 1.00 [4000] |

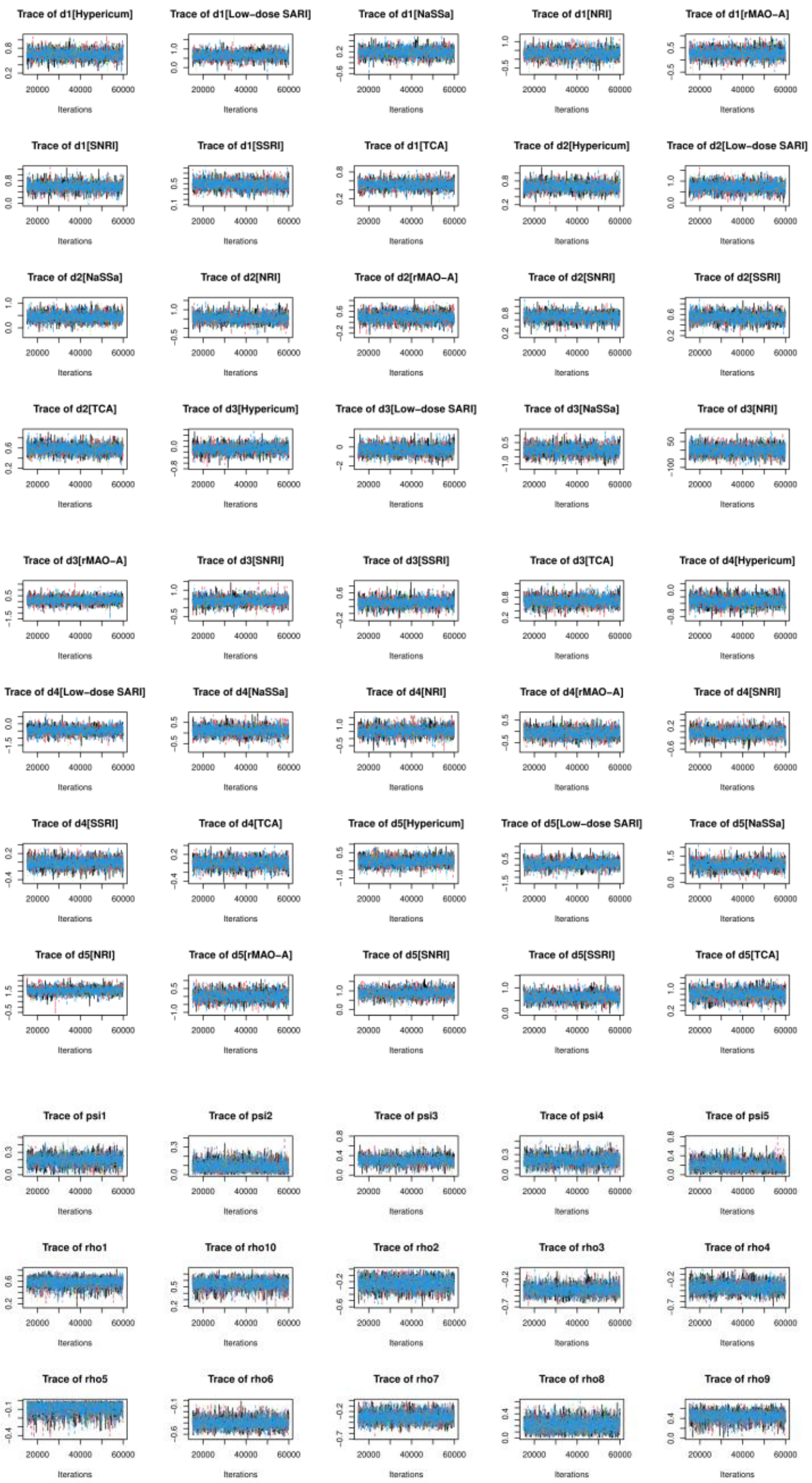

**S-Figure 3:** Traceplots for parameters of interest in the antidepressants dataset for the mvNMA(standard) model. Parameters $d$, $\psi$ (psi) and $\rho$ (rho) correspond to the MCMC chains for basic treatment effects, outcome-specific heterogeneity, and outcome correlation, respectively. The numbers in each trace plot title denote the outcomes, with 1 representing early response, 2 representing early remission, 3 representing adverse events, 4 representing loss to follow-up, and 5 representing loss to follow-up due to adverse events.

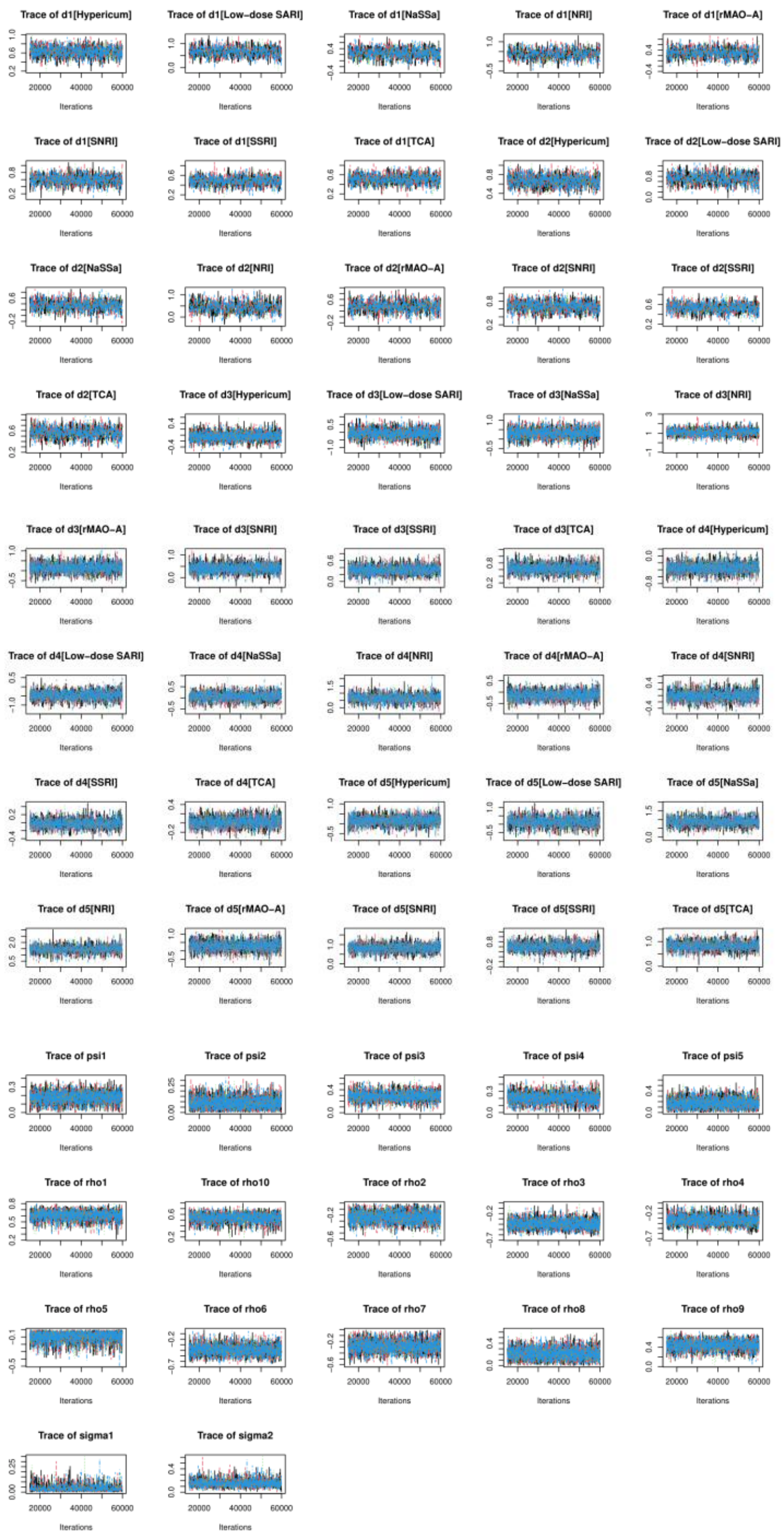

**S-Figure 4:** Traceplots for parameters of interest in the antidepressants dataset for the mvNMA(DM) model. Parameters $d$, $\psi$ (psi) and $\rho$ (rho) correspond to the MCMC chains for basic treatment effects, outcome-specific heterogeneity, and outcome correlation, respectively. The numbers in each trace plot title denote the outcomes, with 1 representing early response, 2 representing early remission, 3 representing adverse events, 4 representing loss to follow-up, and 5 representing loss to follow-up due to adverse events.

### Outcome-specific treatment hierarchies using median ranks

**S-Table 8:** Outcome-specific median ranks and 95% CrIs across all five outcomes of interest, from the mvNMA(standard) model, for the antidepressants dataset.

| Treatment | Early Response | Early Remission | Adverse Events | Loss to follow up | Loss to follow up due to adverse events |
|---|---|---|---|---|---|
| Hypericum | 2 [1, 6] | 3 [1, 7] | 3 [1, 5] | 2 [1, 5] | 2 [1, 4] |
| Low-dose SARI | 2 [1, 7] | 2 [1, 7] | 1 [1, 8] | 1 [1, 6] | 3 [1, 5] |
| SNRI | 3 [1, 7] | 2 [1, 7] | 7 [3, 8] | 5 [2, 8] | 7 [5, 8] |
| TCA | 4 [2, 6] | 4 [2, 7] | 8 [7, 8] | 6 [3, 8] | 6 [5, 8] |
| SSRI | 5 [2, 7] | 5 [3, 7] | 6 [4, 7] | 5 [3, 8] | 5 [4, 6] |
| NRI | 6 [1, 9] | 5 [1, 8] | - | 9 [4, 9] | 9 [8, 9] |
| rMAO-A | 7 [2, 9] | 7 [2, 8] | 4 [1, 7] | 4 [1, 8] | 3 [1, 4] |
| NaSSa | 8 [5, 9] | 7 [2, 8] | 3 [1, 7] | 7 [2, 9] | 8 [5, 9] |
| Placebo | 9 [7, 9] | 9 [8, 9] | 3 [1, 5] | 6 [3, 8] | 2 [1, 4] |

**S-Table 9:** Outcome-specific median ranks and 95% CrIs across all five outcomes of interest, from the mvNMA(DM) model, for the antidepressants dataset.

| Treatment | Early Response | Early Remission | Adverse Events | Loss to follow up | Loss to follow up due to adverse events |
|---|---|---|---|---|---|
| Hypericum | 2 [1, 5] | 2 [1, 6] | 2 [1, 5] | 2 [1, 4] | 3 [1, 4] |
| Low-dose SARI | 2 [1, 7] | 2 [1, 7] | 2 [1, 6] | 1 [1, 5] | 2 [1, 5] |
| SNRI | 3 [1, 6] | 2 [1, 6] | 7 [3, 8] | 5 [2, 8] | 7 [5, 8] |
| TCA | 4 [2, 6] | 4 [2, 7] | 8 [6, 9] | 6 [3, 8] | 7 [6, 8] |
| SSRI | 5 [3, 7] | 5 [3, 7] | 6 [4, 7] | 5 [3, 8] | 5 [4, 6] |
| NRI | 6 [1, 9] | 6 [1, 8] | 9 [7, 9] | 9 [8, 9] | 9 [8, 9] |
| rMAO-A | 7 [3, 9] | 7 [3, 8] | 4 [1, 7] | 3 [1, 8] | 4 [1, 6] |
| NaSSa | 7 [4, 9] | 7 [3, 8] | 5 [2, 8] | 7 [2, 9] | 7 [5, 9] |
| Placebo | 9 [8, 9] | 9 [8, 9] | 3 [1, 5] | 6 [3, 8] | 1 [1, 4] |

## Results of fuzzy VIKOR

**S-Table 10:** Fuzzy VIKOR results for median ranks obtained from the mvNMA(standard) model.

| **Treatment** | **Q** | **S** | **R** |
|---|---|---|---|
| Hypericum | **0.02** | 0.09 | 0.02 |
| Low-dose SARI | **0.04** | 0.09 | 0.03 |
| rMAO-A | **0.32** | 0.17 | 0.05 |
| SSRI | 0.34 | 0.20 | 0.05 |
| SNRI | 0.37 | 0.19 | 0.05 |
| TCA | 0.48 | 0.22 | 0.06 |
| NaSSa | 0.50 | 0.25 | 0.06 |
| Placebo | 0.54 | 0.22 | 0.07 |

**S-Table 11:** Fuzzy VIKOR results for median ranks obtained from the mvNMA(DM) model.

| **Treatment** | **Q** | **S** | **R** |
|---|---|---|---|
| **Low-dose SARI** | **0.03** | 0.07 | 0.02 |
| **Hypericum** | **0.04** | 0.06 | 0.02 |
| SSRI | 0.39 | 0.17 | 0.04 |
| rMAO-A | 0.39 | 0.16 | 0.05 |
| SNRI | 0.41 | 0.16 | 0.05 |
| NaSSa | 0.52 | 0.22 | 0.05 |
| TCA | 0.53 | 0.2 | 0.06 |
| Placebo | 0.58 | 0.19 | 0.06 |
| NRI | 0.7 | 0.26 | 0.06 |

## A re-analysis of the psoriasis data presented by Daly et al.

In their manuscript, Daly et al.[6] presented a network comparing seven biologic therapies for the treatment of psoriasis[7]. Treatments were evaluated across four outcomes, three of which, clear or nearly clear skin, mean change in the Dermatology Life Quality Index (DLQI), and Psoriasis Area and Severity Index (PASI 75), reflect efficacy, while the fourth outcome, withdrawal due to adverse events, captures safety. Outcome-specific SUCRA values obtained from independent NMA models are reported in **S-Table 12**. These results indicate that Ixekizumab performs well across all efficacy outcomes but ranks second lowest with respect to safety.

**S-Table 12:** Outcome-specific treatment rankings using $SUCRA$ ranking metric in the psoriasis data.

| **Treatment** | Clear/nearly clear | Mean change in DLQI | PASI75 | Withdrawal due to adverse events |
|---|---|---|---|---|
| | **SUCRA** | **SUCRA** | **SUCRA** | **SUCRA** |
| Ixekizumab | **0.99** | **0.70** | **0.96** | 0.14 |
| Secukinumab | **0.85** | **0.85** | **0.79** | **0.80** |
| Infliximab | **0.67** | **0.80** | **0.81** | 0.04 |
| Ustekinumab | 0.60 | 0.70 | 0.52 | **0.82** |
| Adalimumab | 0.46 | 0.51 | 0.49 | **0.81** |
| Etanercept | 0.28 | 0.31 | 0.28 | 0.46 |
| Methotrexate | 0.15 | 0.15 | 0.15 | 0.47 |
| Placebo | 0.00 | 0.00 | 0.00 | 0.47 |

Based on the outcome-specific treatment hierarchies reported in **S-Table 12**, we derived an across-outcomes hierarchy using (i) spie charts, following the original publication, and (ii) the proposed VIKOR algorithm. Because only outcome-specific hierarchies were available, and not the full dataset, we could not fully apply the mvNMA-based framework. Nevertheless, we applied VIKOR directly to the outcome-specific hierarchies obtained from independent NMA models, assuming equal outcome weights and a balanced decision-making strategy ($v = 0.5$). Results from both approaches are presented in **S-Table 13**.

Consistent with the original publication[6], spie charts ranked Secukinumab and Ixekizumab as the top two treatments in the across-outcomes hierarchy. This result obscures the fact that Ixekizumab is among the least safe treatments, likely because three efficacy outcomes, where

**S-Table 13:** Treatment ranking across all four outcomes in the psoriasis example. Note that for the VIKOR method, lower values in the respective metrics $Q$, $S$ and $R$ indicate better treatment performance. Treatments identified as the optimal compromise solution by the VIKOR algorithm are shown in bold. The “Treatment” column is ordered according to the $Q$-metric values obtained from the VIKOR approach.

| **Treatment** | **VIKOR (NMA)** | | | **Spie chart (NMA)** |
|---|---|---|---|---|
| | Q | S | R | Area |
| **Secukinumab** | **0.00** | 0.18 | 0.05 | **0.68** |
| Ustekinumab | 0.29 | 0.34 | 0.12 | 0.45 |
| Adalimumab | 0.39 | 0.43 | 0.14 | 0.34 |
| Ixekizumab | 0.50 | 0.30 | 0.22 | 0.60 |
| Infliximab | 0.65 | 0.42 | 0.24 | 0.44 |
| Etanercept | 0.67 | 0.67 | 0.18 | 0.12 |
| Methotrexate | 0.83 | 0.77 | 0.21 | 0.07 |
| Placebo | 1.00 | 0.88 | 0.25 | 0.06 |

Ixekizumab performs well, are combined with a single safety outcome in which it performs poorly. On this basis, Daly et al.[6] concluded that efficacy and safety outcomes should be combined separately when using spie charts.

Results from the VIKOR algorithm (**S-Table 13**) suggest that the method is robust to pronounced conflicts between efficacy and safety outcomes. The combined treatment hierarchy ranks Secukinumab and Ustekinumab at the top, with evaluation of the acceptable advantage and acceptable stability conditions identifying Secukinumab as the unique compromise solution. In contrast, Ixekizumab is now placed in the middle of the across-outcomes hierarchy according to the $Q$ metric. This robustness arises because VIKOR explicitly quantifies and balances overall and worst-case treatment performance when ranking treatments. Notably, if only the overall performance metric $S$ was considered, thereby ignoring the worst-case performance metric $R$, the resulting hierarchy would largely align with the spie chart results and would rank Ixekizumab second.